\documentclass[11pt]{article}
\pdfoutput=1
\usepackage{dcolumn}
\usepackage{bm}

\usepackage{graphicx}
\usepackage{amssymb,amsmath}
\usepackage{multirow}
\usepackage{cite,color,url}
\usepackage[colorlinks=true
,urlcolor=blue
,anchorcolor=blue
,citecolor=blue
,filecolor=blue
,linkcolor=blue
,menucolor=blue
,linktocpage=true
,pdfproducer=medialab
,pdfa=true
]{hyperref}

\usepackage{slashed}
\usepackage{epsfig,psfrag,rotating,soul}
\usepackage{rotfloat}
\usepackage{bbold}

\usepackage[square,numbers,compress,sort]{natbib}
\newcommand{\be}{\begin{equation}}
\newcommand{\ee}{\end{equation}}
\newcommand{\ba}{\begin{eqnarray}}
\newcommand{\ea}{\end{eqnarray}}
\def\Ffactor{f_k}
\def\YLSS{Y_{\rm LSS}}

\evensidemargin \oddsidemargin
\allowdisplaybreaks

\let\OLDthebibliography\thebibliography
\renewcommand\thebibliography[1]{
  \OLDthebibliography{#1}
  \setlength{\parskip}{0pt}
  \setlength{\itemsep}{0pt plus 0.3ex}
}

\begin{document}
\thispagestyle{empty}

\def\thefootnote{\fnsymbol{footnote}}

\begin{flushright}
IFT-UAM/CSIC-19-110\\
FTUAM-19-17\\
LPT-Orsay-19-30
\end{flushright}

\vspace*{1cm}

\begin{center}

\begin{Large}
\textbf{\textsc{Flavor techniques for LFV processes:\\
Higgs decays in a general seesaw model}}
\end{Large}

\vspace{1cm}

{\sc
Xabier Marcano$^{1}$%
\footnote{{\tt \href{mailto:xabier.marcano@th.u-psud.fr}{xabier.marcano@th.u-psud.fr}}}%
~and Roberto A.~Morales$^{2}$%
\footnote{{\tt \href{robertoa.morales@uam.es}{robertoa.morales@uam.es}}}
}

\vspace*{.7cm}

{\sl
$^1$Laboratoire de Physique Th\'eorique, CNRS, \\
Univ. Paris-Sud, Universit\'e Paris-Saclay, 91405 Orsay, France

\vspace*{0.1cm}

$^2$Departamento de F\'{\i}sica Te\'orica and Instituto de F\'{\i}sica Te\'orica, IFT-UAM/CSIC,\\
Universidad Aut\'onoma de Madrid, Cantoblanco, 28049 Madrid, Spain

}

\end{center}

\vspace{0.1cm}

\begin{abstract}
\noindent
Lepton flavor violating processes are optimal observables to test new physics, since they are forbidden in the Standard Model while they may be generated in new theories. 
The usual approach to these processes is to perform the computations in the physical basis; nevertheless this may lose track of the dependence on some of the fundamental parameters, in particular on those at the origin of the flavor violation.
Consequently, in order to obtain analytical expressions directly in terms of these parameters, flavor techniques are often preferred. 
In this work, we focus on the mass insertion approximation technique, which works with the  interaction states instead of the physical ones, and provides diagrammatic expansions of the observables. 
After reviewing the basics of this technique with two simple examples, we apply it to the lepton flavor violating Higgs decays in the framework of a general type-I seesaw model with an arbitrary number of right-handed neutrinos. 
We derive an effective vertex valid to compute these observables when the right-handed neutrino masses are above the electroweak scale and show that we recover previous results obtained for low scale seesaws. 
Finally, we apply current constraints on the model to conclude on maximum Higgs decay rates, which unfortunately are far from current experimental sensitivities. 
\end{abstract}

\def\thefootnote{\arabic{footnote}}
\setcounter{page}{0}
\setcounter{footnote}{0}

\newpage

\tableofcontents

\section{Introduction}
\label{intro}

Lepton flavor violating (LFV) processes are optimal observables to test new physics hypotheses.
This is particularly true for LFV transitions in the charged sector, as
they are forbidden in the Standard Model (SM), and extremely suppressed if one minimally introduces the observed light neutrino masses.
Consequently, any observation of charged LFV transition would be a clear evidence of new physics beyond the SM (BSM).
Moreover, since in several BSM  theories this kind of processes are induced via quantum corrections with new particles running in the loops, exploring LFV transitions in the intensity frontier allows us to probe the existence of these new particles even if they are too heavy to be directly produced in any other experiment. 

Many BSM theories are constructed following the same steps that succeed in the case of the SM. 
First, we write the most general Lagrangian with the chosen symmetries and field content, which includes some parameters that we may consider as the fundamental parameters of the theory.
The easiest way of doing this is by choosing a field basis in which  the conservation of the symmetries, in particular the gauge symmetries, is explicit, implying in most of the cases  that gauge interaction are diagonal in this basis and that some particles are massless. 
Therefore, we  refer to this basis as the interaction basis in general, and also as gauge or flavor basis for the particular cases of gauge or fermionic fields, respectively. 
As a second step, we assume that some of these gauge symmetries are spontaneously broken to a smaller gauge symmetry group,  as in the case of the electroweak symmetry breaking (EWSB), providing masses to some of the fields. 
Nevertheless, these new mass terms do not need to be diagonal in the interaction basis, and some non-diagonal terms may appear, mixing different interaction fields. 
The basis in which the mass terms are diagonal is called the mass or physical basis, since it is in this basis where the parameters could be directly related to observables. 
The relation between the two basis is given by a series of unitary rotation matrices.
In some simple cases, we may be able to find analytical expressions for these rotation matrices in terms of the original parameters in the interaction basis, however this is not always possible and we often need to use numerical methods.

The  physical basis is the natural choice to compute any transition in quantum field theory (QFT), since we can properly define a loop expansion for any observable, such as the LFV processes we mentioned before. 
The reason is that, in this basis, particles have well-defined propagators, meaning that they will keep their identity unless they interact with other two -- or more-- particles. 
The resulting expressions will be given in terms of the physical parameters, i.e., physical masses and rotation matrices.
Nevertheless, in many cases it is desirable to have expressions that, albeit  approximate, are given directly in terms of the fundamental parameters of the interaction basis. 
Indeed, this is particularly interesting in the case of flavor transitions, since we can often track its origin back to few parameters in the original Lagrangian, e.g., the non-diagonal Yukawa couplings in many models. 

One possibility in order to obtain expressions in terms of the fundamental parameters is to compute the amplitudes first in the physical basis, and then expand them using the relations between the two basis. 
An interesting possibility in this direction is given by the Flavor Expansion Theorem~\cite{Dedes:2015twa,Rosiek:2015jua}, which provides a recipe to translate analytically  the expression in the mass basis to the flavor basis. 
Unfortunately, this technique may suffer from limitations when the external momenta and the masses within the loops are such that the involved loop integrals suffer from branch cuts~\cite{Dedes:2015twa}. 
As this will be the case for the observable we will be interested in, we will not consider this technique any further. 
We refer to the original references for more details. 

Alternatively, we will use another technique, the so-called mass insertion approximation (MIA), which is very powerful when computing flavor transitions.
The main idea behind this technique is to perform a new computation, independent to that in the physical basis, working directly in the interaction basis. 
In general, the mass matrices will not be diagonal in this basis and the non-diagonal terms will provide two-point interactions, the so-called mass insertions, which will allow a particle to transform into another one along its propagation.
Then, the propagator of a particle is understood as successive insertions of this kind and, therefore, this technique provides a diagrammatic expansion of any observable. 
Although computing the full series of mass insertions would reproduce the complete result in the physical basis, in general this is not possible to do.
Nevertheless, in the  case where the mass insertions are small -- smaller than the diagonal mass terms --, we can treat these insertions perturbatively and compute the diagrams in the interaction basis to a given order in the MIA expansion. 
For this reason, the MIA technique is very useful to compute LFV processes, since the strong experimental bounds suggest that any kind of parameter leading to LFV transitions should be small. 

\begin{table}[t!]
\begin{center}
\begin{tabular}{lllll}
\hline
\hline
LFV Obs. & \multicolumn{4}{c}{Present Upper Bounds  $(95\%~{\rm C.L.})$} \\
\hline
BR$(H\to\mu e)$ & \hspace{.7cm} - &\hspace{.7cm} & $3.5\times10^{-4}$ &CMS (2016)~\cite{Khachatryan:2016rke}\\
BR$(H\to\tau e)$ & $4.7\times10^{-3}$ &ATLAS (2019)~\cite{Aad:2019ugc} & $6.1\times10^{-3}$ &CMS (2018)~\cite{Sirunyan:2017xzt} \\
BR$(H\to\tau\mu )$ &$2.8\times10^{-3}$ &ATLAS (2019)~\cite{Aad:2019ugc}& $2.5\times10^{-3}$ &CMS (2018)~\cite{Sirunyan:2017xzt}\\
\hline
\hline
\end{tabular}
\caption{Present experimental upper bounds on lepton flavor violating Higgs boson decays. 
Here BR$(H\to \ell_k \ell_m)\equiv \,$BR$(H\to \ell_k \bar{\ell}_m)+$BR$(H\to \bar{\ell}_k \ell_m)$. }
\end{center}
\label{LFVHDexp}
\end{table}

In order to show the applicability of the MIA for computing flavor transitions, we apply it here to the case of LFV Higgs decays (LFVHD). 
The motivation to choose these observables is the strong experimental effort that both ATLAS and CMS are doing in the search for this kind of decays, which we summarize in Table~\ref{LFVHDexp} with the current upper bounds.
Furthermore, the LHCb collaboration has also performed similar searches~\cite{Aaij:2018mea}.
From the theoretical side, these observables have also been studied very actively, exploring their potential to probe BSM theories such as neutrino mass models~\cite{Pilaftsis:1992st,Korner:1992zk,Arganda:2004bz,Arganda:2014dta,Arganda:2015naa,Aoki:2016wyl,Arganda:2017vdb,Herrero-Garcia:2017xdu,Thao:2017qtn},
minimal flavor violation~\cite{Dery:2013aba,Dery:2014kxa,He:2015rqa,Baek:2016pef}, 
supersymmetric models~\cite{Han:2000jz,Curiel:2002pf,DiazCruz:2002er,Curiel:2003uk,Brignole:2003iv, Brignole:2004ah,Parry:2005fp,DiazCruz:2008ry,Crivellin:2010er,Giang:2012vs,Arhrib:2012mg,Arhrib:2012ax,Arana-Catania:2013xma,Arganda:2015uca,Aloni:2015wvn,Vicente:2015cka,Alvarado:2016par,Hammad:2016bng,Gomez:2017dhl}, 
two Higgs doublet models~\cite{Davidson:2010xv,Sierra:2014nqa,Omura:2015nja,Bizot:2015qqo,Botella:2015hoa}, 
composite Higgs models~\cite{Agashe:2009di}, 
models with extra dimensions~\cite{Perez:2008ee,Casagrande:2008hr,Albrecht:2009xr,Azatov:2009na}, 
effective Lagrangians~\cite{DiazCruz:1999xe,deLima:2015pqa,Buschmann:2016uzg,Herrero-Garcia:2016uab,Belusca-Maito:2016axk,Coy:2018bxr} 
and many others~\cite{Blankenburg:2012ex,Harnik:2012pb,Dery:2013rta,Falkowski:2013jya,Campos:2014zaa,Heeck:2014qea,Crivellin:2015mga,Dorsner:2015mja,Baek:2015fma,Huitu:2016pwk,Baek:2016kud,Altmannshofer:2016oaq,DiIura:2016wbx,
Nguyen:2018rlb}.
Indeed, some of these works used  the MIA technique for computing the LFVHD rates~\cite{Arganda:2015uca,Arganda:2017vdb}.

The particular model we will choose to perform the MIA computations will be a general type-I seesaw model~\cite{Minkowski:1977sc,GellMann:1980vs,Yanagida:1979as,Mohapatra:1979ia,Schechter:1980gr}, where an arbitrary number of right-handed (RH) neutrinos are added to the SM. 
This kind of models are very well motivated from the observation of neutrino oscillations~\cite{Fukuda:1998mi,Fukuda:1998ah,Ahmad:2002jz,Eguchi:2002dm}, and it is well-known that they may lead to sizable LFV transitions, see for instance the recent review~\cite{Abada:2018nio} and references therein. 
Furthermore, it has been shown that the MIA works very well in this context~\cite{Arganda:2017vdb,Herrero:2018luu}.

The paper is organized as follows. 
We start by describing the basics for a MIA computation in Section~\ref{MIA}, providing two simple examples with the explicit computations in the case of small and large mass insertions. 
Then, in Section~\ref{GSS} we apply this technique to the case of the LFV Higgs decays in a general seesaw model. 
We use it to derive an effective vertex in the case of heavy RH neutrinos, which helps to compare the results with current experimental bounds from other observables and to conclude on maximum allowed LFVHD rates. 
Finally, we conclude in Section~\ref{sec:conclusions}. 
Further details on the computation and heavy mass expansions are provided in the Appendices.


\section{Basics for a Mass Insertion Approximation computation}
\label{MIA}

In many models, spontaneous symmetry breaking is behind particle masses.
It leads to quadratic terms in the Lagrangian,  which implies a mass matrix in the interaction basis.
In general, this matrix is non-diagonal and when it is diagonalized the physical basis is derived. 
This latter basis is in general the chosen basis to perform QFT computations, as it is possible to properly define a loop expansion for a given observable.
Nevertheless, when doing this one usually loses  track of the parameters in the interaction basis. 
On the other hand, having analytical expressions for a given observable in terms of the fundamental parameters of the theory, it is possible to extract information about them from the experimental measurements directly.

In order to work with the fundamental parameters in the computation of a given observable, the mass insertion approximation provides a powerful tool. This method is a diagrammatic diagonalization of the mass matrix in the interaction basis: in this approach the diagonal entries are considered as the mass parameters, while the non-diagonal ones are interpreted as two-point interactions (the so-called mass insertions) of the corresponding states.
In this context, the propagator of a given state is constructed from the successive mass insertions connecting two different fields and the interaction states are dressed with these consecutive interactions. 
The exact diagonalization corresponds to a complete resummation of the infinite mass insertions that can occur in the propagation. 
In general, it is not possible to do this exact resummation, and therefore an approximation is used: as in a Taylor expansion, a dimensionless parameter is defined as the ratio of the non-diagonal mass insertion over the diagonal mass parameter, and its magnitude defines how many mass insertions must be taken into account to achieve a given precision in the expansion. In a general model, the hierarchy between the different mass scales defines different dimensionless parameters. 

In this Section, we present two examples as an illustration of the application of this technique. 
The first one corresponds to a situation in which  the non-diagonal terms are smaller that the diagonal ones. Then we show that the first two terms in the MIA expansion reproduce the computation in the physical mass basis to that order.
The second example represents the opposite situation: the  non-diagonal parameters are larger than the diagonal ones, and we need to perform a complete resummation of the infinite mass insertions.

\subsection{First example:  small mass insertions}
As a first example of the MIA application, we consider a toy-model composed of three real scalar fields $\rho$, $\Phi_1$ and $\Phi_2$ in the interaction basis, with the following the Lagrangian:
\be
{\cal L}_{\rm gauge} = \frac{1}{2}(\partial_\mu \rho)^2 -\frac{1}{2}\mu_\rho^2\,\rho^2 +\frac{1}{2}(\partial_\mu \Phi_I)^2 -\frac{1}{2} {\bf M}_{IJ}^2\Phi_I\Phi_J -\boldsymbol{\lambda}_{IJ}\,\Phi_I\Phi_J\,\rho\,,
\label{L_toy-model_gauge}
\ee
with an implicit sum over $I,J$ is understood. 

For simplicity, we assume a real and symmetric squared mass matrix but not aligned in the interaction space:
\be
{\bf M}^2 = \left(\begin{array}{cc} M^2 & m^2 \\ m^2 & M^2 \end{array}\right) \,,
\label{massmatrix_toy-model}
\ee
and a cubic interaction between the scalar $\rho$ with the $\Phi$ fields that is diagonal in interaction space:
\be
\boldsymbol{\lambda} = \left(\begin{array}{cc} \lambda & 0 \\ 0 & \lambda \end{array}\right) \,.
\ee
The mass matrix ${\bf M}^2$ can be diagonalized by an orthogonal matrix ${\bf O}$:
\be
\phi_i = {\bf O}_{Ii}\Phi_I \quad\Longrightarrow\quad {\bf O}^T{\bf M}^2{\bf O} = {\rm diag}(M_+^2,M_-^2)\,,
\ee
defining a physical basis $\phi_{\pm}$, with physical masses given by
\be
M_\pm^2 = M^2 \pm m^2 \,.
\label{physmass_toy-model}
\ee
Therefore, the Lagrangian in the physical basis is:
\be
{\cal L}_{\rm phys} = \frac{1}{2}(\partial_\mu \rho)^2 -\frac{1}{2}\mu_\rho^2\,\rho^2 +\frac{1}{2}(\partial_\mu \phi_+)^2 -\frac{1}{2} M_+^2\phi_+^2 +\frac{1}{2}(\partial_\mu \phi_-)^2 -\frac{1}{2} M_-^2\phi_-^2 -\lambda\left(\phi_+^2\rho +\phi_-^2\rho \right)\,.
\label{L_toy-model_phys}
\ee
\begin{figure}[t!]
\begin{center}
\includegraphics[width=0.8\textwidth]{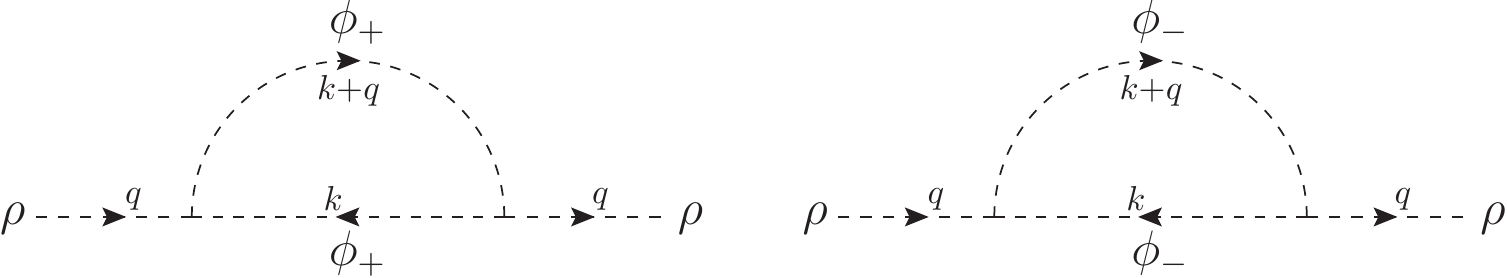}
\caption{One-loop contributions to the self-energy of the scalar $\rho$ in the physical basis, corresponding to the Lagrangian of Eq.~\eqref{L_toy-model_phys}}
\label{diagSE_toy-model_phys}
\end{center}
\end{figure}
Let us consider the one-loop contributions to the self-energy of the $\rho$ scalar field: in the physical basis, they come from the Lagrangian of  Eq.~\eqref{L_toy-model_phys} and correspond to the sunset topology of Fig.~\ref{diagSE_toy-model_phys}, with  $\phi_\pm$ running in the loop.
They are expressed as function of the one-loop integrals defined in App.~\ref{AppA}: 
\ba
-i\Pi(q) &=& \int \frac{d^D k}{(2\pi)^D}(-i\lambda)\frac{i}{k^2-M_+^2}(-i\lambda)\frac{i}{(k+q)^2-M_+^2} \quad+\quad\Big( M_+ \to M_- \Big)  \nonumber\\
&=& \frac{i}{16\pi^2}\lambda^2\Big( B_0(q,M_+,M_+) + B_0(q,M_-,M_-)\Big)\,.
\label{floopSE_toy-model_phys}
\ea
Notice that, in this example, the matrix ${\bf O}$ is not involved in the computation in the physical basis. 
This is due to the simplicity of our model, where we assumed $\boldsymbol{\lambda}=\lambda \mathbb{1}$ and therefore ${\bf O}$ is not present in Eq.~\eqref{L_toy-model_phys}.
In a more general case, the expressions in the physical basis are given terms of the physical masses and the rotation matrices. 

\begin{figure}[t!]
\begin{center}
\includegraphics[width=0.8\textwidth]{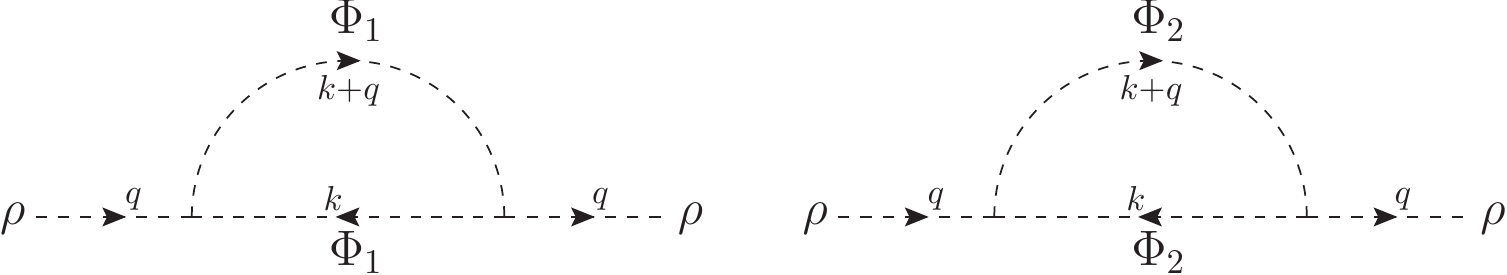}
\includegraphics[width=\textwidth]{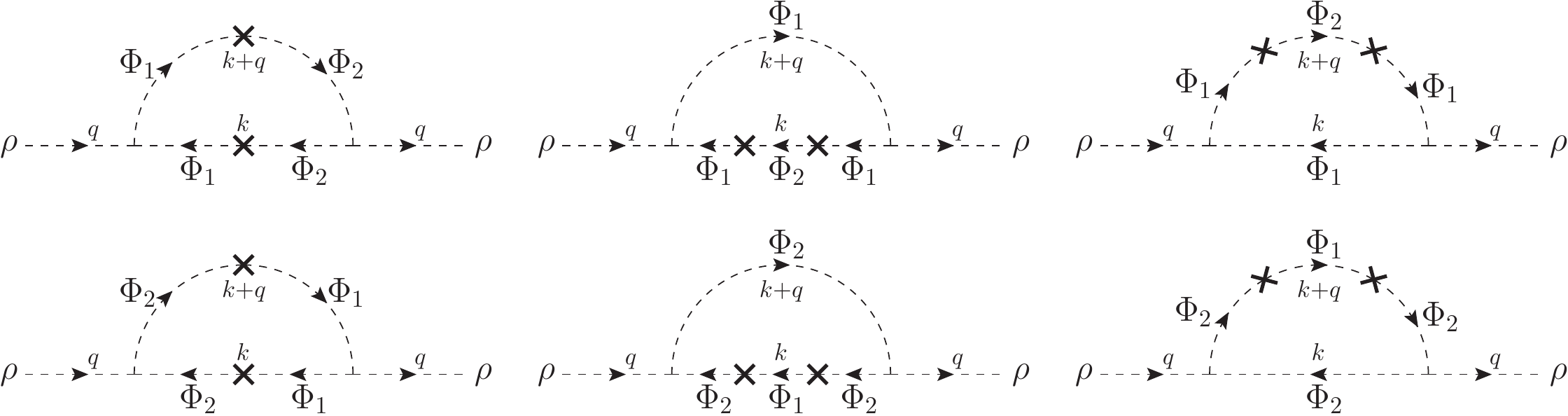}
\caption{One-loop contributions to the self-energy of the scalar $\rho$ in the interaction basis.
The first row corresponds to the LO in the MIA calculation (there is no mass insertion), while the second and third rows define the NLO  (there are two mass insertions denoted by crosses).}
\label{diagSE_toy-model_gauge}
\end{center}
\end{figure}

In order to illustrate how the MIA works, we assume that the non-diagonal entry in the mass matrix ${\bf M}^2$ of Eq.~\eqref{massmatrix_toy-model} is much smaller than the diagonal one. Then the dimensionless parameter $m^2/M^2 \ll 1$ is defined and it controls the diagrammatic expansion in the  interaction basis. 
Now, considering the interaction fields $\Phi_{1,2}$ of Eq.~\eqref{L_toy-model_gauge} running in the loop with an associated mass parameter $M$ and the two-point interaction $\sim m^2\Phi_1\Phi_2$ as the mass insertion, the systematic procedure is to add successive mass insertions up to a given order. 
In  Fig.~\ref{diagSE_toy-model_gauge}, the first two contributions in the MIA are shown: 
they correspond respectively to the leading order (LO) with no mass insertions, and to the next to leading order (NLO) with two insertions.
In this example we cannot close the loop with an odd number of mass insertions in the sunset topology, since the cubic interactions are diagonal.
Notice that all these MIA diagrams are at the same loop level, that of the corresponding diagrams in the physical basis, however they are of different order in the MIA expansion.
Therefore, in this approach we have an expansion for the self-energy at the one-loop level as
\be
-i\Pi^{\rm MIA}(q) = -i\Pi^{\rm MIA}_{\rm LO}(q) -i\Pi^{\rm MIA}_{\rm NLO}(q) \,+\,...
\label{MIAexpansion_toy-model}
\ee
where the dots are contributions with more mass insertions and, thus, suppressed by higher powers of the dimensionless parameter $m^2/M^2$.

The LO contribution in the MIA has the same type of diagrams than in the physical mass basis, but the interaction states $\Phi_{1,2}$ are running in the loop now. Then,
\be
-i\Pi^{\rm MIA}_{\rm LO}(q) = \frac{i}{16\pi^2}2\lambda^2 B_0(q,M,M) \,.
\label{floopSE_toy-model_MIA-LO}
\ee

On the other hand, there are six diagrams contributing to the NLO order in the MIA. Each one has two mass insertion, so they are proportional to $m^4$. Explicitly, the one-loop integral for the middle-left diagram in  Fig.~\ref{diagSE_toy-model_gauge} is given by, 

\ba
&\int \frac{d^D k}{(2\pi)^D}(-i\lambda)\frac{i}{k^2-M^2}\left(-im^2\right)\frac{i}{k^2-M^2}(-i\lambda)\frac{i}{(k+q)^2-M^2}\left(-im^2\right)\frac{i}{(k+q)^2-M^2}  \nonumber\\
&=\frac{i}{16\pi^2}\lambda^2 m^4 D_0(0,q,0,M,M,M,M)\,.
\ea
Then, the NLO contribution in the MIA is:
\begin{align}
-i\Pi^{\rm MIA}_{\rm NLO}(q) &= \frac{i2\lambda^2 m^4}{16\pi^2} \Big[ D_0(0,q,0,M,M,M,M)  +D_0(0,0,q,M,M,M,M) +D_0(q,0,0,M,M,M,M) \Big]\nonumber\\
&= -\frac{i}{16\pi^2}\frac{4\lambda^2 m^4}{q^2-4M^2}\left(\frac{1}{M^2}+2\frac{\log\left(\frac{2M^2-q^2+\sqrt{q^2(q^2-4M^2)}}{2M^2}\right)}{\sqrt{q^2(q^2-4M^2)}}\right)\,.
\label{floopSE_toy-model_MIA-NLO}
\end{align}

The analytical comparison between the MIA and the physical basis results is obtained when the physical masses (and the matrix rotations if any) are expressed in terms of the gauge parameters and the physical basis expressions are expanded up to a given order. 
In this example, since we have used the MIA up to two insertions, we need to expand the expression in the physical basis up to ${\cal O}(m^4/M^4)$. 
From Eqs.~\eqref{physmass_toy-model} and \eqref{floopSE_toy-model_phys}, the physical basis computation of the scalar $\rho$ self-energy in terms of the gauge interaction parameters is
\be
-i\Pi(q) = \frac{i}{16\pi^2}\lambda^2 \bigg( B_0\left(q,\sqrt{M^2+m^2},\sqrt{M^2+m^2}\right) +  B_0\left(q,\sqrt{M^2-m^2},\sqrt{M^2-m^2}\right) \bigg)\,.
\label{floopSE_toy-model_phys2}
\ee
This two-point $B_0$ one-loop function, participating also in the LO contribution of the MIA in Eq.~\eqref{floopSE_toy-model_MIA-LO}, is given by
\be
B_{0} \left(q,w,w\right) = \frac{2}{4-D}-\gamma_E+\log \Big(\frac{4\pi\mu^{2}}{w^2}\Big) +2\, +\frac{\sqrt{q^2(q^2-4w^2)}\log\left(\frac{2w^2-q^2+\sqrt{q^2(q^2-4w^2)}}{2w^2}\right)}{q^2}\,,
\label{B0_toy-model}
\ee
where $\gamma_E$ is the Euler-Mascheroni constant and $\mu$ is the usual scale for dimensional regularization. 

We can now expand the expression obtained in the physical basis, Eq.~\eqref{floopSE_toy-model_phys2}, under the assumption of $m^2\ll M^2$.
At zero order, $m=0$, this equation trivially leads to the LO MIA contribution in Eq.~\eqref{floopSE_toy-model_MIA-LO}.
The next terms in the expansion are of order $m^4/M^4$, and lead to the NLO MIA expression in Eq.~\eqref{floopSE_toy-model_MIA-NLO}.
Similarly, one could check that higher order terms $m^8/M^8, m^{12}/M^{12}, \dots$  will correspond to higher order terms in the MIA expansion with $4, 6,\dots$ insertions.

We remark again that the simplicity of the present toy-model allows us to compare explicitly the MIA results with the expansion of the physical basis results, due to the analytical diagonalization of the $2\times2$ mass matrix.
In a more complex situation, this diagonalization is only performed numerically, and thus the dependence on the interaction parameters is missed. 
Moreover, the computational effort could be huge for higher dimension mass matrices. In that context, the MIA diagrammatic computation is a powerful tool in order to work with the interaction parameters explicitly. As we said, the MIA results correspond to a perturbative calculation in a dimensionless parameter. 

\subsection{Second example:  large mass insertions}\label{Sec:ex2}
\begin{figure}[t!]
\begin{center}
\includegraphics[width=\textwidth]{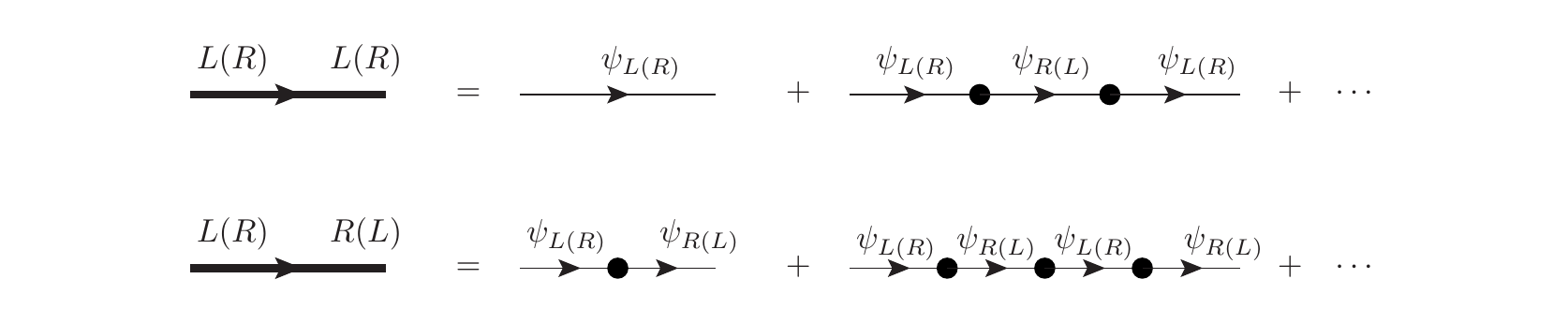}
\caption{Diagrammatic interpretation of the dressed propagators (thick lines) for the same and opposite chiralities as an infinite series of  successive mass insertions (black dots) between two massless propagators (thin lines).}
\label{props_MIA}
\end{center}
\end{figure}

Now we analyze a situation in which a non-diagonal entry of the mass matrix is larger than a diagonal one, i.e., the corresponding dimensionless parameter results larger than 1. In that case, an exact resummation of this large mass insertion is needed. In particular, we consider a Dirac spinor $\psi=P_L\psi+P_R\psi=\psi_L+\psi_R$ with mass $M$ and momentum $p$. The corresponding quadratic terms of the free Dirac Lagrangian are
\ba
{\cal L}_D &=& \bar{\psi}{\slashed p}\psi -M\bar{\psi}\psi \nonumber\\
&=& \overline{\psi_L}{\slashed p}P_L\psi_L +\overline{\psi_R}{\slashed p}P_R\psi_R -\overline{\psi_L}MP_L\psi_L -\overline{\psi_R}MP_R\psi_R  \nonumber\\
&=& \left( \overline{\psi_L} \quad \overline{\psi_R} \right) \left(\begin{array}{cc} {\slashed p}P_L & -MP_R \\ -MP_L & {\slashed p}P_R \end{array}\right) \left(\begin{array}{c} \psi_L \\ \psi_R \end{array}\right) \,,
\label{massmatrix_D}
\ea
where we have a matrix of dimension 2 in  chiral space ($P_{L,R}=(I\mp\gamma_5)/2$). This approach is equivalent to having two massless fermions $\psi_L$ and $\psi_R$ interacting through the mass insertion $M$. 
The corresponding massless propagators are the inverse of the kinetic terms:
\be
{\rm Prop}_{\psi_L} = \frac{i{\slashed p}}{p^2} P_R \quad {\rm and} \quad {\rm Prop}_{\psi_R} = \frac{i{\slashed p}}{p^2} P_L \,.
\ee
As before, the dimensionless parameter of the MIA expansion should be the ratio between the mass insertion $M$ and the mass of the fermions.
However, the latter is zero in this example, implying an infinitely large expansion parameter.
This fact can be solved by defining a dressed propagator that accounts for a resummation of all the insertions of this kind. 
In the chiral basis, there are four types of propagators depending on the chiralities of the connected fermions ($LL$, $LR$, $RL$ and $RR$), as showed in Fig.~\ref{props_MIA}.
Here, the thin lines represent the massless propagators, the blacks dots are the mass insertions and the thick lines correspond to the dressed propagators (after the resummation of the successive two point interactions). The dressed propagators that connect two fermions with the same chirality contain an even number of mass insertions, while the ones connecting two opposite chiralities have an odd number of mass insertions. 

Explicitly, the propagator connecting two left-handed fermions ($LL$) corresponds to the geometric series:
\begin{align}
{\rm Prop}_{\, L\to L} &= \frac{i{\slashed p}}{p^2} P_R + \frac{i{\slashed p}}{p^2}P_R (-i M P_R ) \frac{i{\slashed p}}{p^2} P_L (-i M P_L)\frac{i{\slashed p}}{p^2}  P_R + \cdots \nonumber\\
&=\frac{i{\slashed p}}{p^2} \sum_{n\geq0}\left(\frac{M^2}{p^2}\right)^n P_R  = \frac{i{\slashed p}}{p^2-M^2} P_R\,, 
\label{propLL_MIA}
\end{align}
and the propagator connecting two right-handed fermions ($RR$) is obtained with the interchange $P_L \leftrightarrow P_R$:
\be
{\rm Prop}_{\, R\to R} = \frac{i{\slashed p}}{p^2-M^2} P_L\,. 
\label{propRR_MIA}
\ee

In the same way, the propagator connecting opposite chiralities ($LR$) comes from the geometric series:
\begin{align}
{\rm Prop}_{\, L\to R} &= \frac{i{\slashed p}}{p^2} P_L (-i M P_L) \frac{i{\slashed p}}{p^2} P_R  
+ \frac{i{\slashed p}}{p^2} P_L (-i M P_L)\frac{i{\slashed p}}{p^2}P_R (-i M P_R ) \frac{i{\slashed p}}{p^2} P_L (-i M P_L)\frac{i{\slashed p}}{p^2}  P_R + \cdots 
 \nonumber\\
&=\frac{i M}{p^2} \sum_{n\geq0}\left(\frac{M^2}{p^2}\right)^n P_R = \frac{i M}{p^2-M^2} P_R\,,
\label{propLR_MIA}
\end{align}
and the propagator $RL$ results from the $LR$ after interchanging $P_L \leftrightarrow P_R$:
\be
{\rm Prop}_{\, R\to L} = \frac{i M}{p^2-M^2} P_L\,.
\label{propRL_MIA}
\ee

From the previous relations, we can interpret that the successive non-diagonal two-point interactions dress the propagator providing the corresponding masses.
It is important to connect this approach with four types of propagators with the standard one of the Dirac propagator $\frac{i ({\slashed p}+M)}{p^2-M^2}$: in a generic process with a Dirac propagator, the MIA approach produces four diagrams with the $LL$, $LR$, $RL$ and $RR$ propagators. Adding these contributions from Eqs.~(\ref{propLL_MIA}-\ref{propRL_MIA}), the complete Dirac propagator is restored.
This procedure works in a generic context of two interacting states, as we will see in the next Section for the type-I seesaw model.


\section{MIA in practice: LFV Higgs decays in a general seesaw model}
\label{GSS}

In order to better illustrate the discussion in the previous Section, we apply next the MIA technique to the particular example of LFV Higgs decays in a general type-I seesaw model (GSS), where $N$ right-handed neutrinos are added to the SM. 
The full computation in the neutrino mass basis\footnote{Notice that, in these references,  particular realizations of the type-I seesaw model were considered. Nevertheless, since the expressions are given in the physical basis, the generalization to a GSS is trivially obtained by just changing the range of neutrino indices.} was done in~\cite{Arganda:2004bz} -- see also~\cite{Pilaftsis:1992st} --, and the final expressions after correcting some typos can be found in~\cite{Marcano:2017ucg}. 
The MIA technique was  applied to the particular case of the inverse seesaw model with three pairs of degenerate sterile fermions~\cite{Arganda:2017vdb}.
Here, we generalize these results to a GSS and discuss how to apply them to the particular case of low scale seesaw models, recovering the results of~\cite{Arganda:2017vdb} in the proper limit. 
Finally, we apply the constraints from the global fit analysis in~\cite{Fernandez-Martinez:2016lgt} to conclude on the maximum allowed $H\to\ell_k\ell_m$ rates. 

\subsection{Model setup for the MIA}

We consider a general type-I seesaw model where the SM is extended with $N$ right-handed neutrinos. The corresponding Lagrangian is given by
\begin{equation}\label{LagGSS}
\mathcal L_{\rm GSS} = -Y_\nu^{ia}\, \overline{L_i} \widetilde\Phi \nu_{R_a} - \frac12 M^{ab}\, \overline{\nu_{R_a}^{\,c}} \nu_{R_b}^{} + h.c.
\end{equation}
where $L$ is the SM left-handed lepton doublet and $\widetilde \Phi= i\sigma_2 \Phi^*$ with $\Phi$ the SM Higgs doublet. The fundamental parameters of the model are then the neutrino Yukawa coupling $Y_\nu$, which is a $3\times N$ complex matrix, and the Majorana mass matrix $M$ which is a $N\times N$ symmetric matrix that violates lepton number in two units.
The $C$-conjugate is defined as usual as $\psi^c=\mathcal C \bar\psi^T$, where we can choose $\mathcal C = i \gamma_2\gamma_0$.
After the EWSB, this Lagrangian leads to a neutrino mass matrix that, in the flavor basis $(\nu_L^{\,c}, \nu_R)$, reads
\begin{equation}\label{MatGSS}
M_{\rm GSS}=\left(\begin{array}{cc} 0 & m_D \\ m_D^T & M \end{array}\right)\,,
\end{equation}
with the Dirac mass matrix defined as $m_D=v Y_\nu$ and $v\simeq174$~GeV.
In the seesaw limit, the non-diagonal entries $m_D$ are smaller than the diagonal $M$ and, therefore we can perform a MIA computation, which will be defined as a perturbative expansion in powers of $m_D/M$.

Moreover, and despite the fact we will be interested in expressing our results in terms of the EW parameters $m_D$ and $M$, we recall that in this seesaw limit the physical masses of the heavy neutrino will be approximately given by the Majorana mass matrix $M$, and that the active-sterile mixings in the physical basis will be of the order $m_D/M$. 
Thus, our MIA computation will be in this sense an expansion in terms of active-sterile mixings.

As discussed in the previous Section, the first step in a MIA computation is choosing the proper basis. 
Despite the fact that the MIA works in the flavor basis, it is not mandatory to work with the full model in this basis, and it is actually  convenient to choose the basis for each sector independently. 
For the present exercise of computing the LFV Higgs decays at one-loop, we will choose a hybrid basis: we will work with the flavor basis for the neutrino sector, while the rest of the fields will be taken in their mass basis. 
The latter  will apply to the external fields, Higgs boson and charged leptons, as well as to the gauge and Goldstone bosons in the loops, which we will treat in the Feynman-'t Hooft gauge\footnote{A full computation in a general $R_\xi$ gauge and proof of gauge invariance can be found in~\cite{Arganda:2017vdb}. }. 
By doing this, we will obtain a useful expression for the LFVHD rates in terms of the new flavor parameters in Eq.~\eqref{LagGSS} and those already known SM parameters in the physical basis. 

Moreover, in the following we will choose the $\nu_R$ basis such that the Majorana mass matrix is real and diagonal. 
Notice that we can do this without any loss of generality, and it will only imply that, for models where $M$ is not diagonal at the beginning, we will need to diagonalize it first, as we will do explicitly when discussing low scale seesaw models in Section~\ref{LowSS}. 
If $M$ is diagonal, then the only lepton flavor changing mass insertion will be the Dirac mass term and, consequently, our MIA computation will be defined as an expansion of successive mass insertions of $m_D$. 
Nevertheless, in the computation of our observable there is another source of LFV coming from the $Y_\nu$ of the cubic interactions between the $\nu_R$ with the L doublet and the $H$ and Goldstone bosons, see the first term of Eq.~\eqref{LagGSS}.
Since the actual source of LFV is the same in both interaction and mass insertion, it will be convenient to consider the Yukawa coupling as the relevant LFV parameter for the expansion and organize our contributions in powers of the $Y_\nu$, as we will see later.

\begin{figure}[t!]
\begin{center}\includegraphics[scale=1]{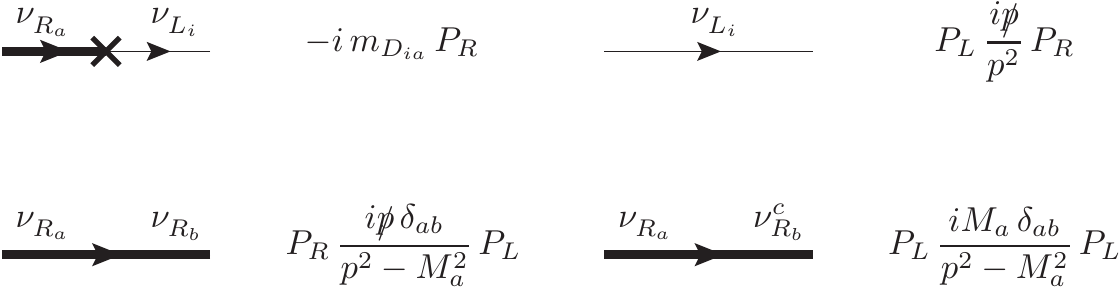}
\caption{Relevant (dressed) propagators and LFV mass insertion for a general seesaw. 
All the other Feynman rules needed for the computation of $H\to \ell_k \bar\ell_m$ at one-loop can be found in~\cite{Arganda:2017vdb}.}
\label{GSS_FR}
\end{center}
\end{figure}

Once we have chosen to work in the flavor basis for the neutrino sector, the Majorana mass matrix $M_{\rm GSS}$ can be understood as the collection of all the relevant mass insertions.
From the one side, the already mentioned $m_D$ will mix the $\nu_L$ and $\nu_R$ fields and we will denote it with a cross, as in Fig.~\ref{GSS_FR}.
Since every insertion of this kind will introduce a new $\nu_R$ field, each new insertion is expected to be suppressed by inverse powers of the heavy mass $M$ and, therefore, we can treat the $m_D$ mass insertions perturbatively. 
On the other hand, the mass term $M$ can be understood as a -- lepton number violating -- mass insertion between the $\nu_R$ and $\nu_R^{\, c}$ fields. 
This mass insertion, however, is not small and, thus, we need to resum all possible insertions of this kind. 
Following the discussion in the previous Section~\ref{Sec:ex2}, we can define two kinds of dressed propagators, a lepton number conserving (LNC) one with any even number of insertions, and a lepton number violating (LNV) one with an odd number of insertions:
\begin{align}
{\rm Prop}_{\, \nu_{R_a}^{}\to\nu_{R_a}^{}}
&
= P_R~~~ \dfrac i{\slashed p}~\,\sum_{n\geq0}\bigg(\dfrac{M_{a}^2}{p^2}\bigg)^n ~ P_L
= P_R~ \dfrac{ i\slashed p}{p^2-M_{a}^2}~ P_L\,, \label{propLNC}
\\
{\rm Prop}_{\, \nu_{R_a}^{}\to \nu_{R_a}^{\, c}}
&
= P_L~ \dfrac {i M_{a}}{p^2}\sum_{n\geq0}\bigg(\dfrac{M_{a}^2}{p^2}\bigg)^n ~ P_L
= P_L~ \dfrac{ i M_{a}}{p^2-M_{a}^2}~ P_L\,,\label{propLNV}
\end{align}
where $M_a$ is the corresponding element of the diagonal mass matrix $M$.
Notice that we are interested in computing a lepton number conserving process; hence, it will be enough to consider the first of these dressed propagators. 

With this setup, the computation is basically the same than that performed in~\cite{Arganda:2017vdb}, with the dressed propagator in Eq.~\eqref{propLNC} playing the role of the {\it fat-propagator} in~\cite{Arganda:2017vdb}.
All the other Feynman rules relevant for the computation of the LFVHD are the same, so we refer to~\cite{Arganda:2017vdb} for further details and conventions.


\subsection{The MIA computation and the heavy mass expansions}
\label{MIA-GSS}

\begin{figure}[t!]
\begin{center}
\includegraphics[width=.9\textwidth]{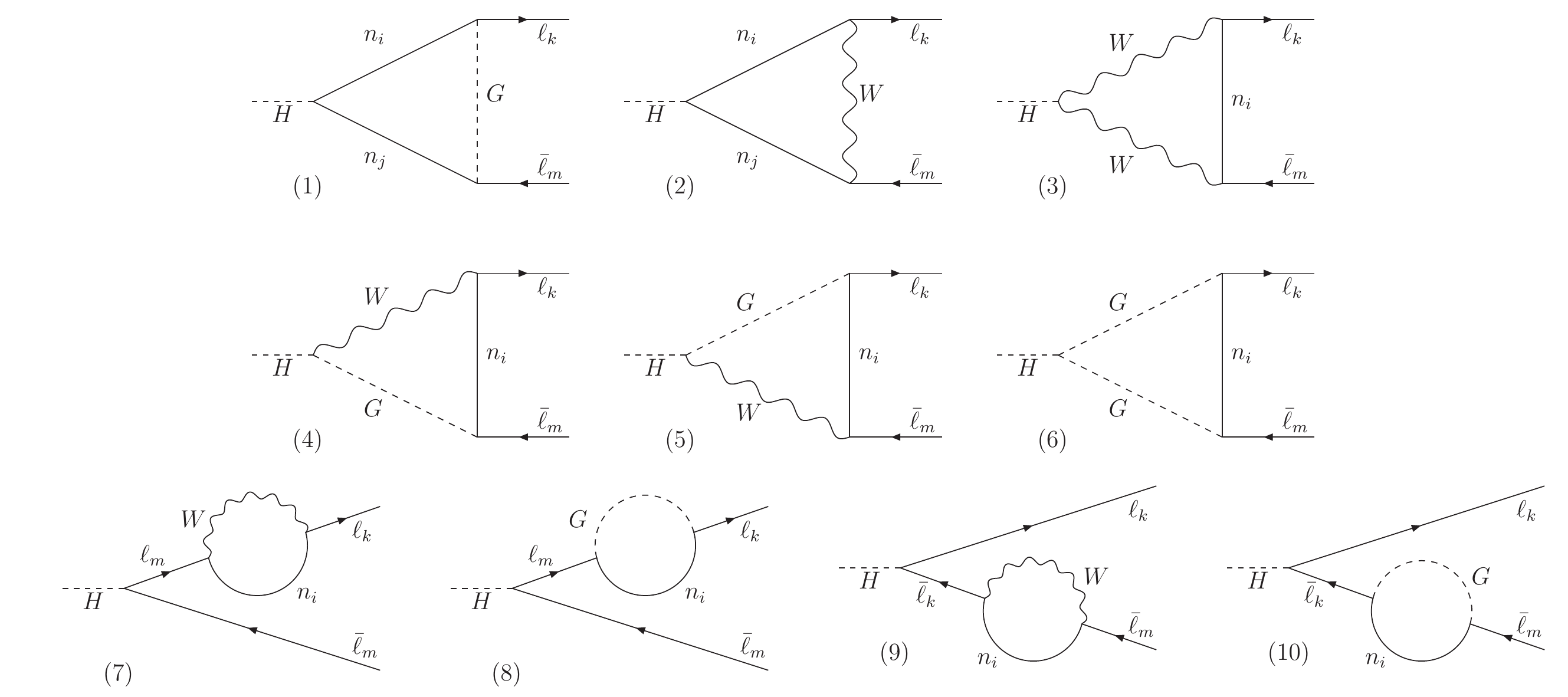}
\caption{One-loop diagrams in the Feynman-'t Hooft gauge for the process $H\to\ell_k {\bar \ell_m}$  in the neutrino mass basis.}
\label{DiagramsMassbasis}
\end{center}
\end{figure}

We are interested in computing the LFV process $H(p_1) \to \ell_k(-p_2) \bar \ell_m (p_3)$, whose decay amplitude can be generically decomposed in terms of two form factors $F_L$ and $F_R$~\cite{Arganda:2004bz},
\begin{equation}
i {\cal M} = -i g \bar{u}_{\ell_k} (-p_2) (F_L P_L + F_R P_R) v_{\ell_m}(p_3) \,,
\label{amplitude}
\end{equation}
In order to further simplify our expressions, one could neglect  the masses of the charged leptons with respect to the Higgs boson mass. 
Nevertheless, the form factor $F_{L(R)}$ is proportional to $m_{\ell_{k(m)}}$, so we cannot fully neglect lepton masses and we need to keep the leading term. 
Using the fact that charged lepton masses are hierarchical, we work under the hypothesis $m_{\ell_m}\ll m_{\ell_k}$. Then, it is enough to consider the $F_L$ form factor for the $H\to \ell_k\bar\ell_m$ decay, keeping the leading contribution in $m_{\ell_k}$ and neglecting any additional contribution from charged lepton masses\footnote{For complete results, we refer to~\cite{Arganda:2017vdb,Marcano:2017ucg}.}. 
Then, the partial decay width can be written as:
\begin{equation}\label{decaywidth}
\Gamma (H \to {\ell_k} \bar{\ell}_m) = \frac{g^2}{16 \pi} m_H |F_L|^2\,.
\end{equation}
Equivalently, in this case where $m_{\ell_m}\ll m_{\ell_k}$, $F_R$ dominates in the CP-conjugated process $H\to \ell_m\bar\ell_k$.
In the Feynman-'t Hooft gauge and in the neutrino mass basis, these form factors receive contributions from the diagrams in Fig.~\ref{DiagramsMassbasis},
\begin{equation}
F_{L,R}=\sum_{{\rm i}=1}^{10} F_{L,R}^{\rm (i)}\,.
\label{FFLR}
\end{equation}
We remark that the definition of mass basis depends on the perturbation order considered, as loop corrections will generically modified the mass matrix adding non-diagonal terms, which need to be rotated away.
We chose to work with the tree-level mass basis and, consequently, we need to include the self-energy corrections to the external legs, last line in Fig.~\ref{DiagramsMassbasis}.
Alternatively, one can work with the one-loop level mass basis, which would rotate away those diagrams. 
Nevertheless, both techniques are equivalent at the one-loop level. 

The full analytical expressions for these form factors can be found in~\cite{Marcano:2017ucg}. 
These expressions are given in terms of the physical neutrino masses and the unitary matrix that relates the flavor and physical basis, so the analytical dependence on the initial parameters in Eq.~\eqref{LagGSS} is lost.
Morover, evaluating numerically these expression could be time consuming, in particular when the amount of right-handed neutrinos is large. 
Therefore, it would be useful to have expressions which are given directly in terms of the flavor parameters in Eq.~\eqref{LagGSS}, even if they are approximations.

As we already know the full result in the physical basis, we could in principle apply the Flavor Expansion Theorem proposed in~\cite{Dedes:2015twa}. 
Nevertheless, this technique requires that the external momenta are smaller than the masses running in the loops, and this is not the case in the decay process $H\to\ell_k\bar\ell_m$ due to the fact that the external momentum of the on-shell Higgs is larger than the $\nu_L$ and $W$ masses.
Therefore, we do not consider this technique for this computation. We perform instead a MIA expansion, since it can be applied even when light masses are running in the loops. 

In the MIA, this process can be computed as an expansion of the relevant LFV mass insertions, which in our case are the Dirac mass terms, or equivalently the Yukawa couplings.
Thus, each of the diagrams in Fig.~\ref{DiagramsMassbasis} can be computed in the MIA as,
\begin{equation}
F_{L,R}^{{\rm MIA\, (i)}} = F_{L,R}^{{\rm MIA\, (i)}} \big|_{Y_\nu^2} + F_{L,R}^{{\rm MIA\, (i)}} \big|_{Y_\nu^4} + \dots
\label{FFMIA}
\end{equation}
The fact of having only even powers of $Y_\nu$ is related to the right-handed neutrinos, whose presence is needed to induce the LFV transition. 
Since they only interact via the Yukawa couplings, each (dressed) RH propagator will introduce two $Y_\nu$, one at each edge of the propagator. 
This means that the LO terms $\mathcal O(Y_\nu^2)$ will come from MIA diagrams with only one RH propagator, the NLO corrections $\mathcal O(Y_\nu^4)$ from diagrams with two, and so on. 
Being the Yukawa coupling perturbative, it should be enough to compute the first contributions to this expansion.
Moreover, the addition of RH propagators will introduce inverse powers of $M$ and ensure the convergence of  Eq.~\eqref{FFMIA} as a perturbative expansion in terms of $\mathcal O(m_D^2/M^2)$.

We collect the MIA results in App.~\ref{AppA}, as well as the relevant Feynman diagrams entering the computation. 
We include the complete $\mathcal O(Y_\nu^2)$ terms, which give a good description of the observable when the Yukawa couplings are small. 
Nevertheless, for large Yukawa couplings, we need to compute also some of the $\mathcal O(Y_\nu^4)$ terms~\cite{Arganda:2017vdb}, which are not as suppressed as we may naively expect from the above discussion.
These dominant $\mathcal O(Y_\nu^4)$ terms are also given in App.~\ref{AppA}.

In order to better understand this point, it is useful to analyze our results when the Majorana scale is heavier than the electroweak scale. 
Indeed, the $O(Y_\nu^4)$ terms may become relevant when the Yukawa couplings are large and, since we are working under the hypothesis $m_D\ll M$, it implies heavy Majorana masses.  
Under this hypothesis of heavy $M$, the loop integrals contributing to the form factors in App.~\ref{AppA}  can be expanded in inverse powers of $M$, as shown in App.~\ref{AppB}. 
The obtained result for the form factor can be interpreted as a low-energy effective vertex induced from heavy Majorana neutrinos, $F_L\equiv V_{H\ell_k\bar\ell_m}^{\rm eff}$.

In this heavy Majorana mass limit, the $\mathcal O(Y_\nu^2)$ terms in the MIA expansion contribute dominantly to the order $\mathcal O(m_D^2/M^2)$, as expected. 
Similarly, we might naively expect that the  $\mathcal O(Y_\nu^4)$ will contribute as $\mathcal O(m_D^4/M^4)$ and will be, therefore,  negligible. 
Nevertheless, it turns out that some diagrams lead to $\mathcal O(Y_\nu^4)$ terms that are suppressed only by two inverse powers of $M$, and thus they are important to take into account for a good MIA prediction. 
Indeed, for very large couplings $\mathcal O(Y_\nu)\sim 1$, these terms may become the dominant ones. 
In the particular process we are considering, they come from diagrams of type (1), (8) and (10) in Fig.~\ref{DiagramsMassbasis}, whose dominant MIA $\mathcal O(Y_\nu^4)$ diagrams are shown in Fig.~\ref{dominant-Y4} of  App.~\ref{AppB}.

It is interesting to discuss a bit more the presence of this kind of {\it unsuppressed} $\mathcal O(Y_\nu^4)$ terms in different LFV observables induced from heavy neutrinos. 
Besides in the LFV Higgs decays, similar contributions were found in the context of LFV Z boson decays~\cite{Herrero:2018luu}, which at the same time suggests that they are also present in LFV 3-body decays, such as $\ell_k\to 3\ell_m$, due to the strong correlation between these two latter observables~\cite{Abada:2014cca,DeRomeri:2016gum}.
On the contrary, these terms are not present in the case of LFV radiative decays $\ell_k\to \ell_m\gamma$~\cite{Arganda:2014dta}. 
The difference could be tracked to the fact that neutrinos do not couple to the photon, but they do couple to the $H$ and $Z$ bosons, leading for example to diagram (1) in Fig.~\ref{DiagramsMassbasis} for the LFVHD, which is not present in the $\ell_k\to\ell_m\gamma$ process\footnote{In the neutrino mass basis, this can be understood as the additional contribution due to the neutrino neutral current, as discussed in e.g.~\cite{Pilaftsis:1992st,Abada:2015zea}.}.
The fact that the radiative decays are different for these other LFV processes is very interesting, as the former are usually the most constraining LFV processes, however this may not be true at very large Yukawa couplings due to these additional terms.

Now, collecting all the relevant terms of $\mathcal O(M^{-2})$, we arrive to the following effective vertex for the LFV $H\to \ell_k\bar\ell_m$ decay,
\ba
V_{H\ell_{k}\bar\ell_{m}}^{\rm eff} &=& \frac{1}{64 \pi^{2}} \frac{m_{\ell_k}}{m_{W}}  \left[  \frac{m_{H}^{2}}{M_{a}^{2}}
\left( r\Big(\frac{m_{W}^{2}}{m_{H}^{2}}\Big) +\log\left(\frac{m_{W}^{2}}{M_{a}^{2}}\right) \right) \left(Y_{\nu}^{ka} Y_{\nu}^{\dagger am}\right)  \right.  \nonumber\\
&&\left. \hspace{19mm}-2v^2 {\cal G}(M_a,M_b) \left(Y_{\nu}^{ka} (Y_{\nu}^{\dagger} Y_{\nu})^{ab} Y_{\nu}^{\dagger bm} \right)  \right]\,,
\label{VeffGSS}
\ea
where we have defined:
\be
r(\lambda)=-\frac{1}{2} -\lambda -8\lambda^{2} +2(1-2\lambda +8\lambda^{2})\sqrt{4\lambda-1}\arctan\left(\frac{1}{\sqrt{4\lambda-1}}\right) +16\lambda^{2}(1-2\lambda)\arctan^2\left(\frac{1}{\sqrt{4\lambda-1}}\right)  \,,\nonumber\\ 
\ee 

\be
{\cal G}(x,y) = \frac{x^2-y^2+(x^2-2y^2)\log \left(\frac{x^{2}}{y^{2}}\right)}{(x^2-y^2)^2}\,,
\qquad 
{\cal G}(x,x) = \frac{3}{2x^2}\,.
\ee
For the physical values of $m_H=125$~GeV and $m_W=80.4$~GeV we have $r(m_W^2/m_H^2)\sim 0.31$.
We recall again that this expression is valid under the assumption of heavy Majorana masses $M\gg v$ and in the seesaw limit $m_D\ll M$, since we  only kept the $\mathcal O(M^{-2})$ terms and we performed the computation at NLO in the MIA expansion. 
As we will see later, these terms are enough to reproduce the full results to a good accuracy in the parameter space that is still allowed by current constraints.

In Eq.~\eqref{VeffGSS}, we have to sum over the indices $a$ and $b$, which run over the RH neutrinos.
In general, all of them will contribute and the indices will run from 1 to $N$. Nevertheless, in some interesting cases some of the RH neutrinos might be very heavy and they will completely decouple from the observable.
Since the contribution to any very heavy neutrino to Eq.~\eqref{VeffGSS} is negligible, decoupling a RH neutrino is indeed equivalent to restricting the range of $a$ and $b$ to those (non-decoupled) right-handed neutrinos, which are still light enough to contribute. 

Another interesting limit corresponds to the case with complete degenerate RH neutrinos, i.e., $M_1=...=M_N\equiv M$. 
In this case, the effective vertex corresponds to
\be
V_{H\ell_{k}\bar\ell_{m}}^{\rm eff} = \frac{1}{64 \pi^{2}} \frac{m_{\ell_k}}{m_{W}}  \left[  \frac{m_{H}^{2}}{M^{2}}
\left( r\Big(\frac{m_{W}^{2}}{m_{H}^{2}}\Big) +\log\left(\frac{m_{W}^{2}}{M^{2}}\right) \right) Y_{\nu} Y_{\nu}^{\dagger} -\frac{3v^2}{M^2} Y_{\nu} Y_{\nu}^{\dagger} Y_{\nu} Y_{\nu}^{\dagger}  \right]^{km}\,.
\label{VeffGSSdeg}
\ee

Notice that, even if have focused on the $H\to \ell_k\bar\ell_m$ channel, the effective vertex $V_{H\ell_{m}\bar\ell_{k}}^{\rm eff}$ of the CP-conjugated process $H\to \ell_m\bar\ell_k$ can be easily obtained by conjugating   Eq.~\eqref{VeffGSS}:
\be
V_{H\ell_{m}\bar\ell_{k}}^{\rm eff} = \left(V_{H\ell_{k}\bar\ell_{m}}^{\rm eff}\right)^*\,,
\ee
which is equivalent to interchanging the flavor index of the charged leptons $k$ and $m$ in the Yukawa couplings.

Finally, the branching ratio for the process $V_{H\ell_{k}\bar\ell_{m}}^{\rm eff}$ can be computed by just plugging the corresponding effective vertex in  Eq.~\eqref{decaywidth}, 
\begin{equation}\label{BReff}
{\rm BR} (H \to {\ell_k} \bar{\ell}_m) = \frac{g^2}{16 \pi} \frac{m_H}{\Gamma_H^{\rm tot}} |V_{H\ell_{k}\bar\ell_{m}}^{\rm eff}|^2\,,
\end{equation}
where $\Gamma_H^{\rm tot}$ is the total width of the Higgs boson.

\begin{figure}[t!]
\begin{center}
\includegraphics[width=.49\textwidth]{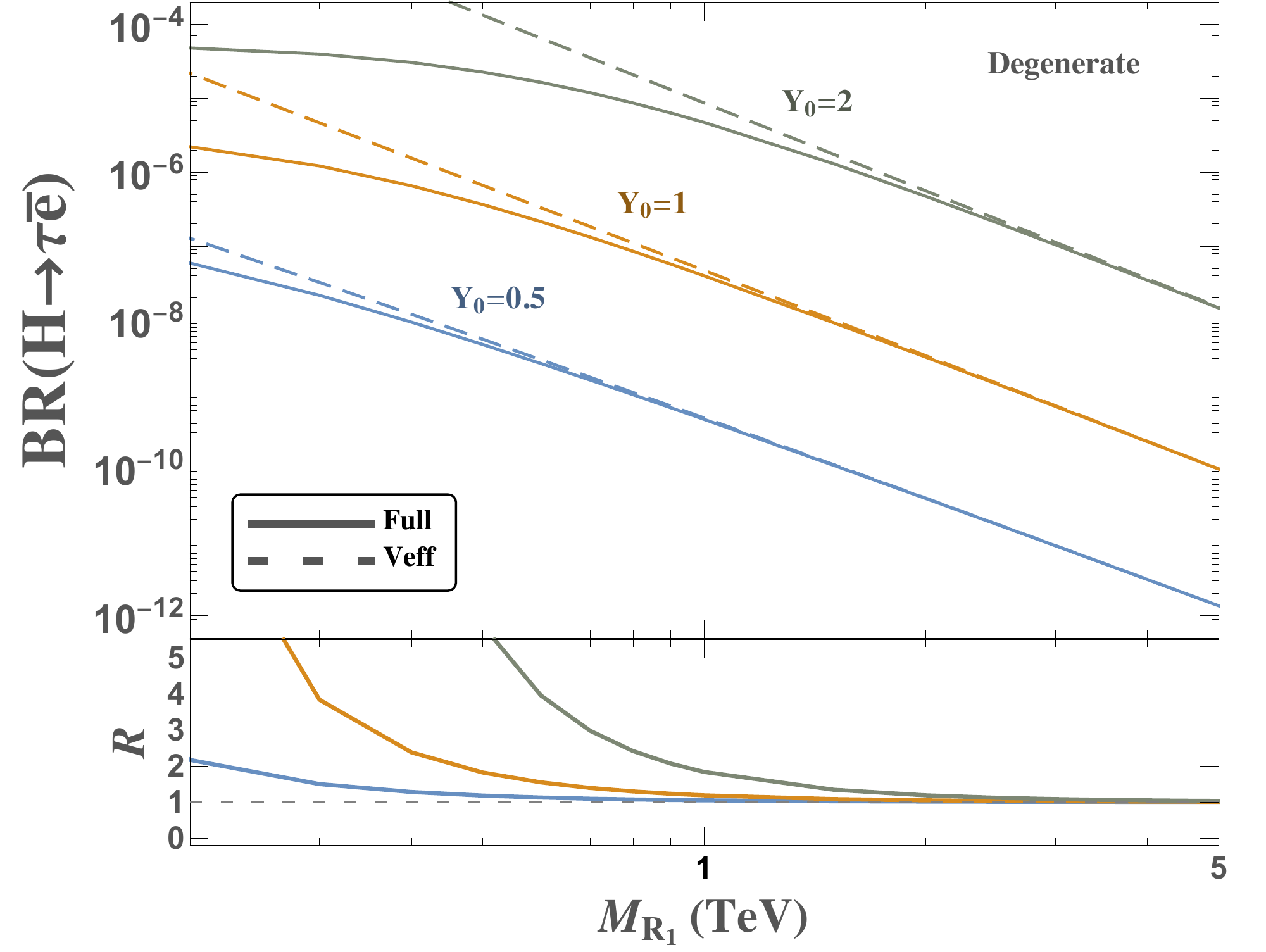}
\includegraphics[width=.49\textwidth]{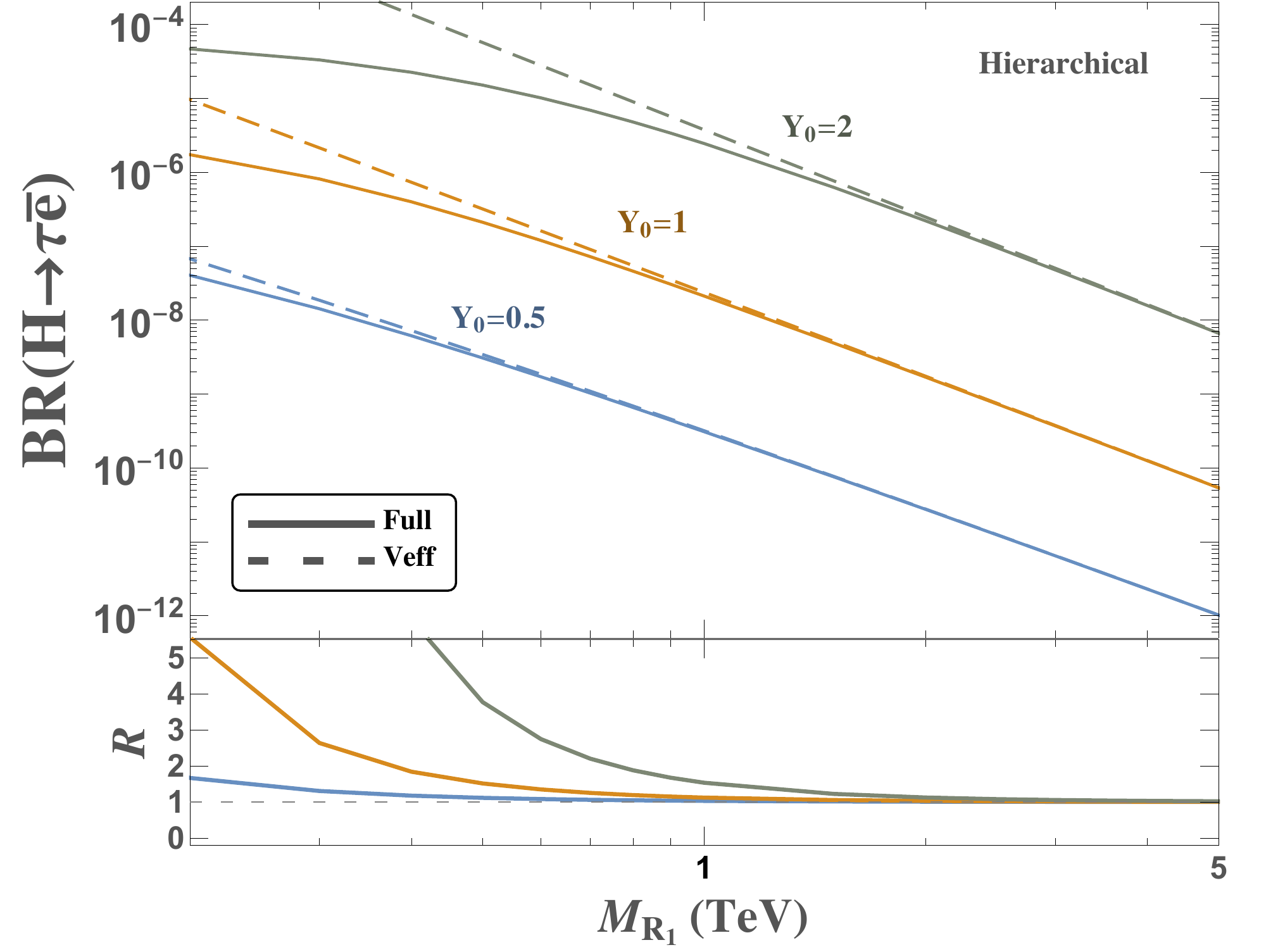}
\caption{Prediction for BR($H\to\tau \bar e$) using the exact computation (solid lines) and the effective vertex in Eq.~\eqref{VeffGSS} (dashed lines). 
The ratio $R={\rm BR}_{V_{\rm eff}}/{\rm BR}_{\rm Full}$ quantifies the agreement between both predictions.
The particular seesaw scenario is defined in Eq.~\eqref{YLSSexample}, with $Y_0$ controlling the global strength of the Yukawa coupling. 
We consider degenerated heavy neutrinos in the left ($M_{R_2}=M_{R_1}$) and hierarchical in the right ($M_{R_2}=10M_{R_1}$).}\label{LFVHD_Rratio}
\end{center}
\end{figure}

In order to illustrate the accuracy of the effective vertex, we show in Fig.~\ref{LFVHD_Rratio} the predictions for $H\to\tau \bar e$ computed using the full expressions in the mass basis (solid lines) as well as using the approximated expression in Eq.~\eqref{BReff}.
Moreover, we quantify the agreement by means of the ratio $R$, defined as the approximated prediction over the full one.
We choose here a particular realization of the seesaw model that we will introduce in Eq.~\eqref{YLSSexample}, although we found similar results for other examples. 
The differences between the two panels are the heavy neutrino masses, chosen to be degenerated in the left and hierarchical in the right.
The overall conclusion from this figure is that the effective vertex in Eq.~\eqref{VeffGSS} works very well in both cases, as long as the condition $m_D\ll M$ is fulfilled. 


\subsection{Connection to low scale seesaw models}
\label{LowSS}

As a particular but interesting application of the effective vertex in Eq.~\eqref{VeffGSS}, we apply it to the so-called low scale seesaw models (LSS), such as the inverse~\cite{Mohapatra:1986aw,Mohapatra:1986bd,Bernabeu:1987gr} or linear~\cite{Malinsky:2005bi} seesaw models. 
These models are of great phenomenological interest since they can introduce relatively light heavy neutrinos with large Yukawa couplings, and still accommodate naturally the observed light neutrino masses.
Moreover, this will also allow us to compare with existing results in the inverse seesaw model~\cite{Arganda:2017vdb}.

The common feature in all these models is the imposition of an approximate conservation of lepton number~\cite{Moffat:2017feq}, which is only violated by some small parameter related to light neutrino masses. 
The difference between each particular low scale seesaw realization is precisely the nature or origin of this small LNV parameter. 
However all of them share the same lepton number conserving limit. 
In our case of study, the LFVHD are LNC processes, and thus the small LNV parameters will not play any important role and we can be neglected.
This means that we can expect to have the same LFVHD rates for all these low scale seesaw models. 

The LNC low scale seesaws could be realized by adding $n_\nu$ pairs of new fermionic singlets with opposite lepton number, which we denote $\nu_R^a$ and $\nu_S^a$ respectively, with $a=1,\dots,n_\nu$. 
For the purpose of this discussion, we are just interested in the neutrino mass matrix\footnote{For more details on LSS, see for instance~\cite{Weiland:2013wha}.}, which in the $(\nu_L^{\,c}, \nu_R, \nu_S)$ basis reads as 
\begin{equation}
M_{\rm LSS} = \left(\begin{array}{ccc}
0 & v\YLSS & 0 \\
v\YLSS^T & 0 & M_R \\
0 & M_R^T & 0 
\end{array}\right)\,,
\label{MassLSS}
\end{equation}
where $\YLSS$ and $M_R$ are respectively $3\times n_\nu$ and $n_\nu\times n_\nu$ matrices, and the zeros have the proper dimensions so the total $M_{\rm LSS}$ matrix is a $(3+2n_\nu)$ symmetric matrix.

In order to apply our results from the previous Section, the first step is to rotate the heavy neutrino sector to its diagonal form,
\begin{equation}
U^T M_{\rm LSS}\, U = \left(\begin{array}{cc}
0 & m_D \\ m_D^T & M_{\rm diag}
\end{array}\right)\,
\end{equation}
where now the dimensions of the new Dirac $m_D$ and Majorana mass $M_{\rm diag}$ matrices are $3\times 2n_\nu$ and $2n_\nu\times2n_\nu$, respectively, and $U$ is a unitary matrix rotating the neutrino sector.
Without any loss of generality, we can assume that $M_R$ is already diagonal and, therefore, the diagonalization becomes trivial, 
\begin{equation}\label{Mdiag-LSS}
U = \left(\begin{array}{cc} \mathbb{1}_{{}_ 3} & 0 \\ 0 & V \end{array}\right)\,,
\quad{\rm with}~~
M_{\rm diag} = V^T\,\left(\begin{array}{cc} 0 & M_R \\ M_R & 0 \end{array}\right)\, V
=\left(\begin{array}{cc}  M_R & 0\\ 0& M_R \end{array}\right)\,,
\end{equation}
where the unitary matrix $V$ just contains  rotations of $\pi/4$ and $i$ factors to make the entries of $M_{\rm diag}$ possitive. 
In general, this unitary rotation is defined by four blocks of $n_\nu\times n_\nu$
\begin{equation}
V = 1/\sqrt2 \left(\begin{array}{cc}
\mathbb1_{n_\nu} &- i\, \mathbb1_{n_\nu} \\
\mathbb1_{n_\nu} & \phantom{-}i\, \mathbb1_{n_\nu} 
\end{array}\right)\,.
\end{equation}

Finally, the new Dirac mass matrix is given by
\begin{equation}\label{MD-LSS}
m_D = \big(v\YLSS ~~0 \big)\, V = \frac{v}{\sqrt2} \big( \YLSS ~~ -i\, \YLSS \big)\,.
\end{equation}

We have now all the pieces needed to compute the $H\to\ell_k\bar\ell_m$ process, we just need to plug the Dirac matrix $m_D=v Y_\nu$ of Eq.~\eqref{MD-LSS} and the $M_{\rm diag}$ of Eq.~\eqref{Mdiag-LSS}, in the effective vertex of Eq.~\eqref{VeffGSS}.
For instance, we can consider the same setup as in~\cite{Arganda:2017vdb}, where all the entries of $M_R$ are degenerate.  
In that case, we can use the Eq.~\eqref{VeffGSSdeg}, with
\begin{equation}
m_D\, m_D^\dagger = \frac{v^2}{2} \big( \YLSS~~ -i\; \YLSS)\, \left(\begin{array}{c} \YLSS^\dagger\\ i\, \YLSS^\dagger\end{array}\right)
= v^2\,\YLSS^{\phantom{\dagger}}\, \YLSS^\dagger\,,
\label{GSStoLSS}
\end{equation}
resulting in the effective vertex
\be
V_{H\ell_{k}\bar\ell_{m}}^{\rm eff} = \frac{1}{64 \pi^{2}} \frac{m_{\ell_k}}{m_{W}}  \left[  \frac{m_{H}^{2}}{M_R^{2}}
\left( r\Big(\frac{m_{W}^{2}}{m_{H}^{2}}\Big) +\log\left(\frac{m_{W}^{2}}{M_R^{2}}\right) \right) \YLSS^{\phantom{\dagger}} \YLSS^{\dagger}
-\frac{3v^2}{M_R^2} \YLSS^{\phantom{\dagger}} \YLSS^{\dagger} \YLSS^{\phantom{\dagger}} \YLSS^{\dagger}  \right]^{km},
\label{VeffISS}
\ee
which is agreement with the result in~\cite{Arganda:2017vdb}, obtained for the particular case of the inverse seesaw model.
Notice that, even that this equation seems to be the same as Eq.~\eqref{VeffGSSdeg}, it is now expressed in terms of the parameters of the low scale seesaw parameters in Eq.~\eqref{MassLSS}, whose physical interpretation is different from the parameters in Eq.~\eqref{MatGSS}.

\subsection{Numerical analysis of the LFV Higgs decays}
\label{numerical}

We conclude this Section by applying the derived effective vertex to study how large the LFVHD rates could be in a GSS model after having considered possible constraints from other observables. 
For that purpose, we will follow the global fit analysis done in~\cite{Fernandez-Martinez:2016lgt}, where two different scenarios were considered: a model with only 3 heavy RH neutrinos (3N-SS), and  a general seesaw with an arbitrary number of them, as in Eq.~\eqref{LagGSS}. 
In both case, the Authors obtained upper bounds on the $\eta$ matrix, a small Hermitian matrix  encoding the deviations from unitarity in the light neutrino mixing. 
In our case, this matrix can be expressed as\footnote{Notice that our definition of $Y_\nu^{}$  corresponds to $Y_\nu^\dagger$ in~\cite{Fernandez-Martinez:2016lgt}.},
\begin{equation}
\eta = \frac12 m_D^{} M^{-2} m_D^\dagger = \frac12 v^2 Y_\nu^{} M^{-2} Y_\nu^\dagger\,,
\end{equation}
and, at the $2\sigma$ level, it is bounded to be below\footnote{These bounds correspond actually to the GSS, although very similar bounds are obtained for the 3N-SS scenario.}
\begin{equation}\label{etamax}
\big|2\eta_{km}\big| \leq 10^{-3} \cdot
\left(\begin{array}{ccc}
2.5 & 0.024 & 2.7 \\
0.024 & 0.4 & 1.2 \\
2.7 & 1.2 & 5.6
\end{array}\right)\,.
\end{equation}

It is interesting to analyze first the 3N-SS case, as it is simpler. 
If we assume again the LNC limit, then it can be implemented by 
\begin{equation}\label{3NSS}
Y_\nu^{\text{3N-SS}} = \left( \begin{array}{ccc} Y_e & 0 &0 \\ Y_\mu &0&0\\ Y_\tau &0 &0\end{array}\right)\,,
\qquad
M^{\text{3N-SS}} = \left(\begin{array}{ccc} 0 & \Lambda &0 \\ \Lambda &0&0 \\ 0&0&\Lambda'\end{array}\right)\,.
\end{equation}
Notice that in this LNC limit light neutrinos are strictly massless, but they can be accommodated by introducing small LNV parameters in these matrices~\cite{Fernandez-Martinez:2016lgt}. 
Nevertheless, since the LFVHD do not violate lepton number, these small LNV parameters will not be relevant for our observable and, therefore, we neglect them in the following.

In this scenario, the effective vertex in Eq.~\eqref{VeffGSS} becomes,
\begin{equation}\label{Veff-3NSS}
V_{H\ell_{k}\bar\ell_{m}}^{\text{\scriptsize 3N-SS}}=\frac1{64\pi^2} \frac{m_{\ell_k}}{m_{W}}
\, Y_k^{}Y^*_m\,
\left\{\frac{m_H^2}{\Lambda^2} \left( r\Big(\frac{m_{W}^{2}}{m_{H}^{2}}\Big) +\log\Big(\frac{m_{W}^{2}}{\Lambda^{2}}\Big)\right)
- \frac{3v^2}{\Lambda^2} \Big(|Y_e|^2 +|Y_\mu|^2+|Y_\tau|^2\Big) \right\}\,,
\end{equation}
where we have used that $Y_\nu Y_\nu^\dagger Y_\nu Y_\nu^\dagger= \left(|Y_e|^2 +|Y_\mu|^2+|Y_\tau|^2\right) Y_\nu Y_\nu^\dagger$.
Alternatively, it can be also written in terms of $\eta$ as, 
\begin{equation}\label{Veff-3NSSeta}
V_{H\ell_{k}\bar\ell_{m}}^{\text{\scriptsize 3N-SS}}=\frac1{64\pi^2} \frac{m_{\ell_k}}{m_{W}}
\, (2\eta_{km})\,
\left\{\frac{m_H^2}{v^2} \left( r\Big(\frac{m_{W}^{2}}{m_{H}^{2}}\Big) +\log\Big(\frac{m_{W}^{2}}{\Lambda^{2}}\Big)\right)
- \frac{3\Lambda^2}{v^2} \Big(|2\eta_{ee}| +|2\eta_{\mu\mu}|+|2\eta_{\tau\tau}|\Big) \right\}\,.
\end{equation}
Notice that the observable vanishes in the $\Lambda\to\infty$ limit, as it is manifest in Eq.~\eqref{Veff-3NSS}. 
This decoupling behavior is hidden when we express it in terms of $\eta$, but this form is useful to conclude on maximum allowed LFVHD rates in this model. 
Due to the $\mathcal O(Y_\nu^4)$ terms, the maximum rates will be obtained at the largest value of $\Lambda$ that allows to saturate Eq.~\eqref{etamax} without spoiling the perturbativity of the Yukawa couplings.
Assuming a perturbativity bound of $|(Y_\nu Y_\nu^\dagger)^{km}|<4\pi$, a rough estimation points to $\Lambda \approx 10$~TeV and consequently,
\begin{align}
&{\rm BR}(H\to \mu e) \lesssim 10^{-14}\,,  \\
&{\rm BR}(H\to \tau e) \lesssim 10^{-8}\,, \\
&{\rm BR}(H\to \tau\mu) \lesssim 10^{-9}\,,
\end{align}
where we have defined BR$(H\to \ell_k\ell_m)\equiv$ BR$(H\to \ell_k\bar\ell_m)$+BR$(H\to \ell_m\bar\ell_k)$. 
The differences between the three channels have a double origin.
From the one side, the fact that these decays are proportional to charged lepton masses suppressed the $H\to \mu e$ with respect to the other two by a factor of $m_\mu^2/m_\tau^2$.
On the other hand, current bounds in Eq.~\eqref{etamax} are a bit stronger for $\eta_{\tau\mu}$, and much more stringent for $\eta_{\mu e}$ due to the strong constraints coming from $\mu\to e\gamma$~\cite{TheMEG:2016wtm}.
This double suppression makes of $H\to\mu e$ an even more challenging observable for experiments.

\begin{figure}[t!]
\begin{center}
\includegraphics[width=.49\textwidth]{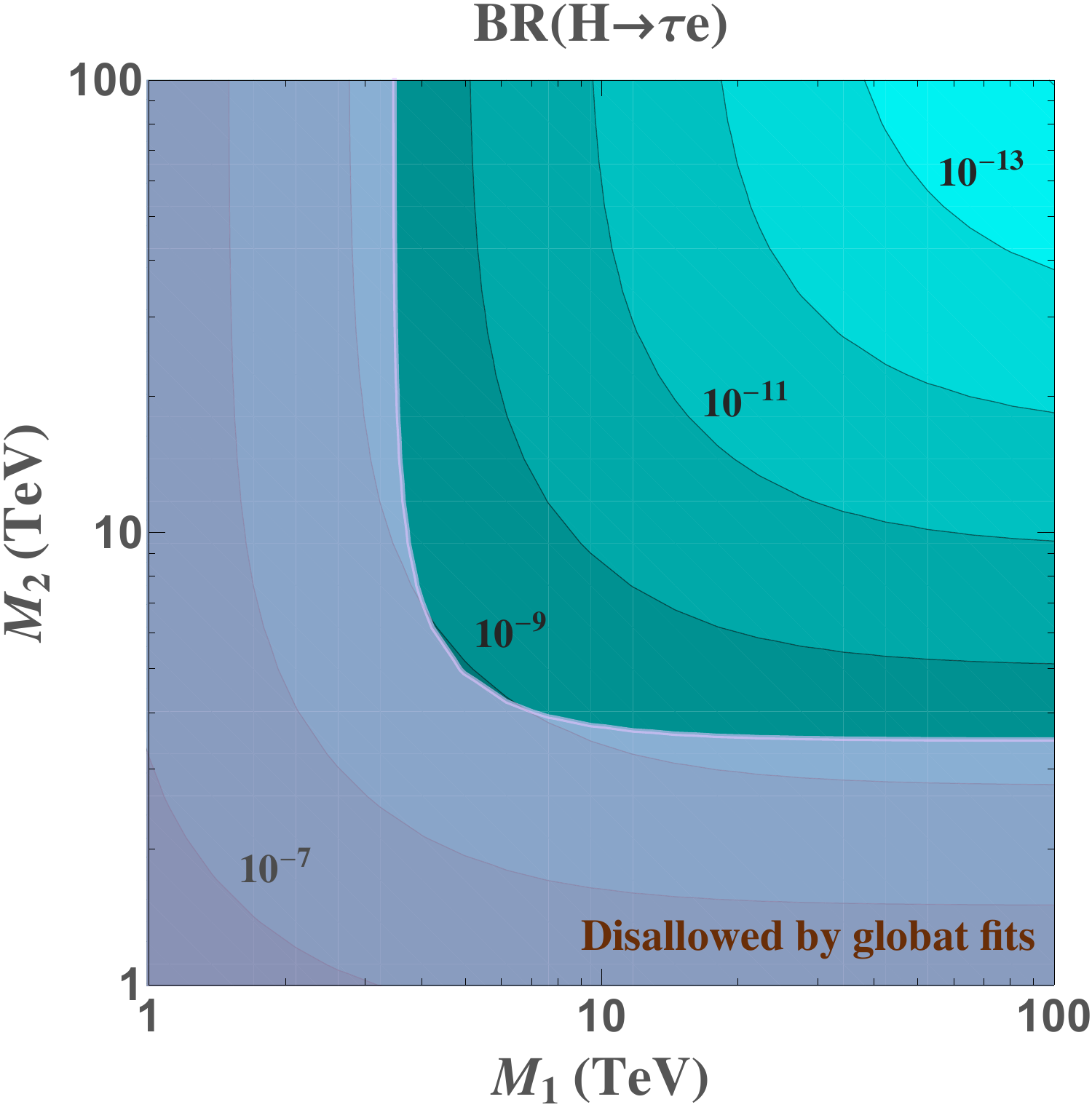}
\includegraphics[width=.49\textwidth]{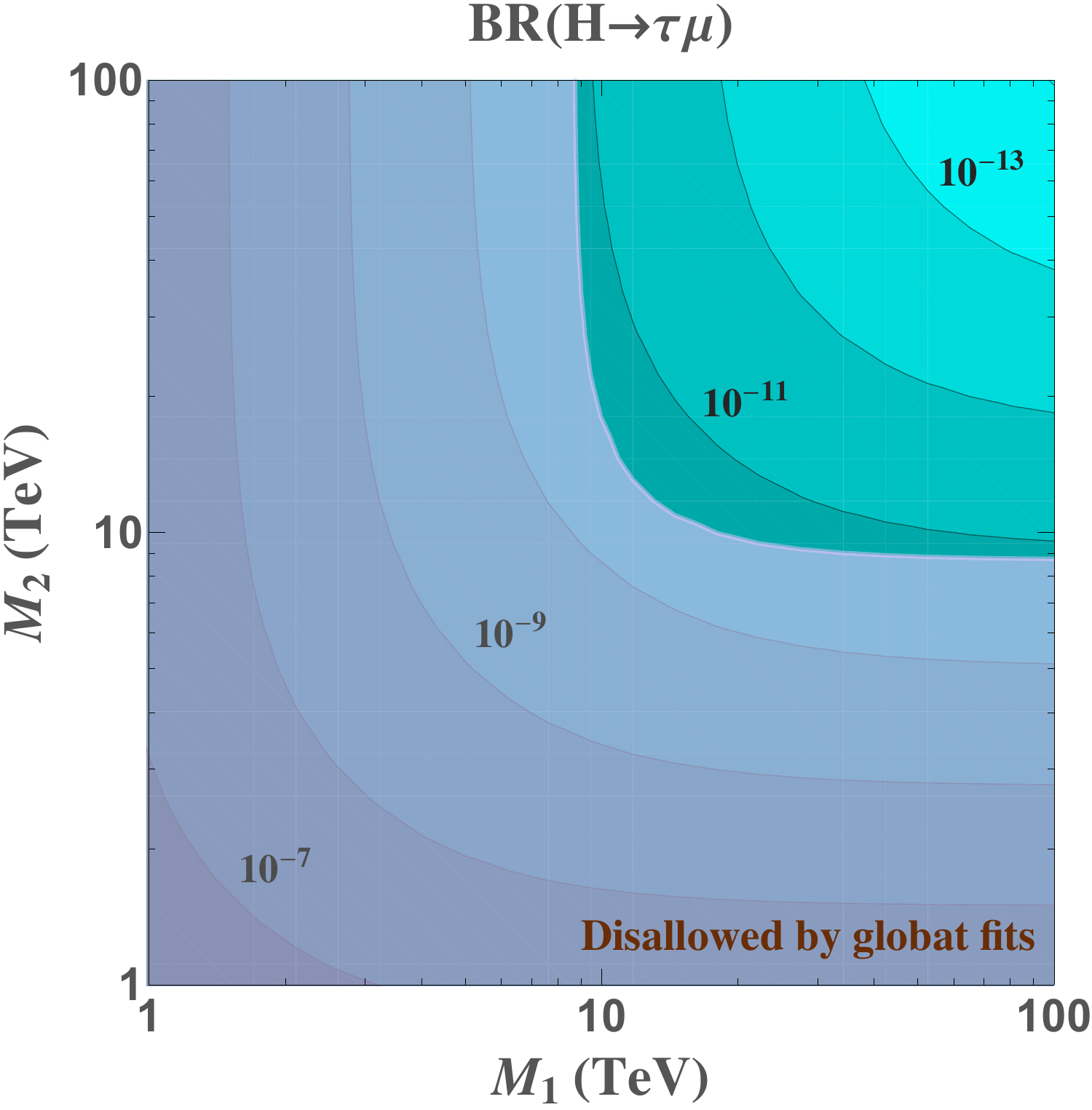}
\caption{Predictions for  LFV Higgs boson decays using Eq.~\eqref{VeffGSS} in a seesaw model with two pairs of Majorana neutrinos with masses $M_{R_1}$ and $M_{R_2}$, as defined in Eq.~\eqref{MassLSS}. 
In the left, both neutrinos have Yukawa couplings equal to 1 with electron and tau leptons, and 0 with muons.
In the right, we exchange the roles of electrons and muons.   
Purple shadowed area is disallowed by global fit constraints in Eq.~\eqref{etamax}.
}
\label{LFVHD_contours}
\end{center}
\end{figure}

The effective vertex in Eq.~\eqref{VeffGSS} has the advantage of being more general than previously computed ones.
In particular, it allows to explore scenarios where the heavy neutrinos have different Majorana masses, as we show in Fig.~\ref{LFVHD_contours}.
In order not to generate too large masses for light neutrinos, we consider an scenario such as the one in Eq.~\eqref{MassLSS}, but with only two pairs of Majorana neutrinos ($n_\nu=2$) contributing to the Higgs decays, assuming that any possible additional neutrino has decoupled from this observable.  
Moreover, and in order to avoid the strong bounds in the $\mu$-$e$ sector, we have considered a simplified case for the Yukawa couplings where none of the heavy neutrinos couple to muons (left panel) or to electrons (right panel). 
The rest of the entries of the Yukawa coupling matrix are set to one, again for simplicity.  
This figure has been done using Eq.~\eqref{VeffGSS}, although we have checked that the full computation leads to the same results in the relevant area allowed by the constraints of Eq.~\eqref{etamax}.

In this Fig.~\ref{LFVHD_contours} we find again the decoupling behavior with each of the individual heavy masses, as we discussed before. 
We can also see that the predictions for $H\to\tau e$ and $H\to\tau\mu$ are the same in both panels, as we have assumed a simplified scenario with equal size Yukawa couplings to the relevant flavors. 
Nevertheless, the bounds of Eq.~\eqref{etamax} are stronger in the muon sector than in the electron one, which again implies that the allowed rates are larger for $H\to\tau e$ than for $H\to\tau\mu$.
Finally, we notice that the results are symmetric with respect to $M_{R_1}$ and $M_{R_2}$, although this is again a consequence of our simplified hypothesis of equal Yukawa couplings, and in general this will not be the case.

In a more general scenario, the $\eta$ matrix will constrain a combination of all the heavy neutrino mixing to the active state, each of these contributions being of order $\xi_{\ell a}=m_D^{\ell a}/M_{a}$, where ${\xi=m_D M^{-1}}$ is the usual seesaw parameter. 
Generalizing the description of~\cite{Arganda:2017vdb} to the case of having $N$ RH neutrinos, we can think of this parameter as made of three $N$-vectors:
\begin{equation}
\xi=\left(\begin{array}{c} \xi_e \\ \xi_\mu \\ \xi_\tau \end{array}\right) =
\left(\begin{array}{cccc}
\xi_{e1} & \xi_{e2} & \cdots & \xi_{eN} \\
\xi_{\mu1} & \xi_{\mu2} & \cdots & \xi_{\mu N} \\
\xi_{\tau1} & \xi_{\tau2} & \cdots & \xi_{\tau N} 
\end{array}\right)\,,
\end{equation}
which leads to
\begin{equation}\label{YLSS}
2\eta = \xi \xi^\dagger = \left(\begin{array}{ccc}
|\xi_e|^2 & \xi_e \cdot\xi_{\mu}^* & \xi_e \cdot \xi_\tau^* \\
\xi_\mu\cdot\xi_e^* & |\xi_\mu|^2 & \xi_\mu\cdot \xi_\tau^* \\
\xi_\tau\cdot \xi_e^* & \xi_\tau\cdot \xi_\mu^* & |\xi_\tau|^2
\end{array}\right)\,.
\end{equation}
This implies that the diagonal entries of the $\eta$ matrix constrain the modulus of these $N$-vectors, while the diagonal ones set upper bounds on the complex scalar products between them. 
Moreover, this geometrical picture is also useful to find solutions that accommodate the current bounds given in Eq.~\eqref{etamax}, as we only need to set the modulus and angles between these vectors, and to explore the implications on our LFV observable. 

\begin{figure}[t!]
\begin{center}
\includegraphics[width=.49\textwidth]{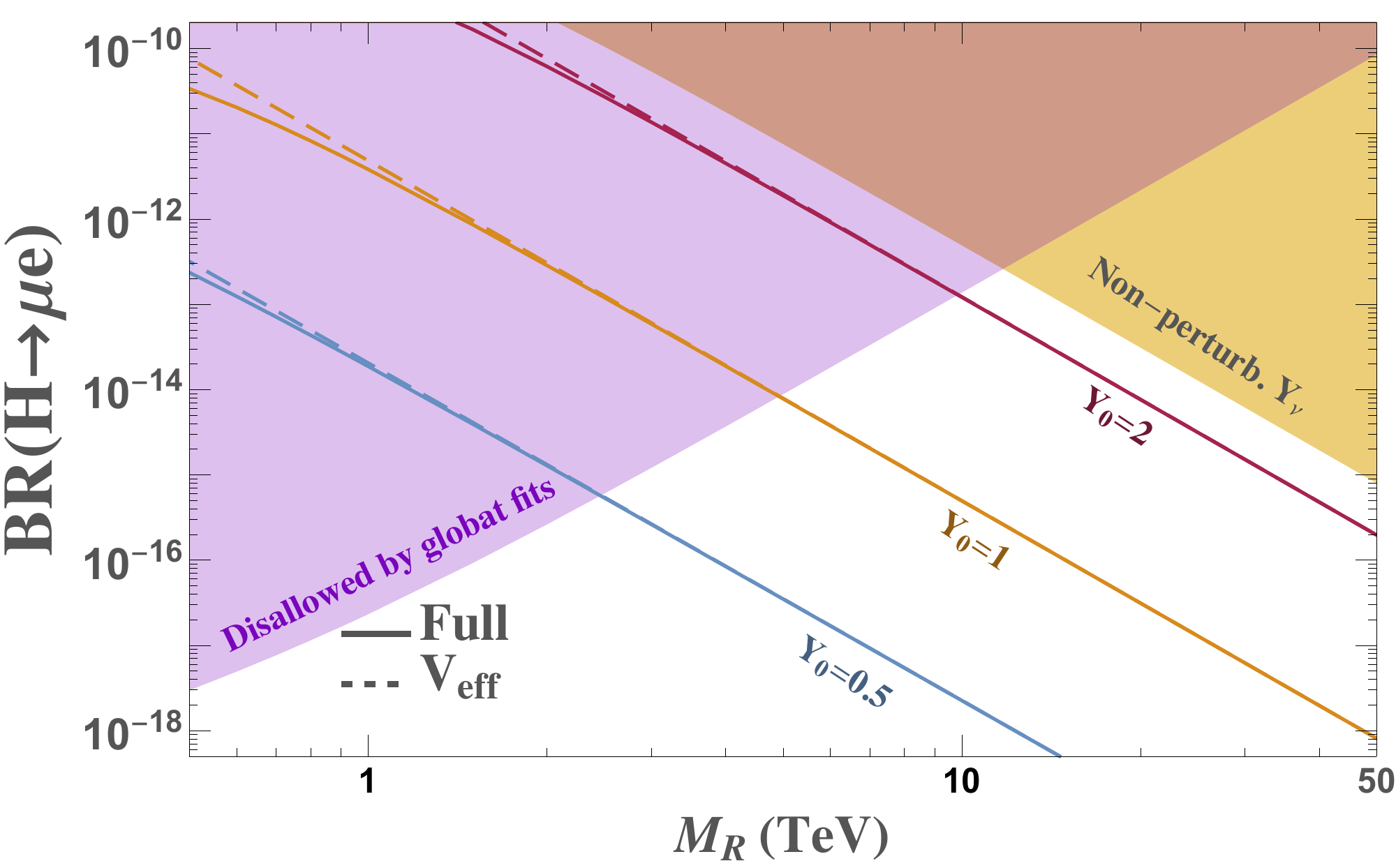}\\
\includegraphics[width=.49\textwidth]{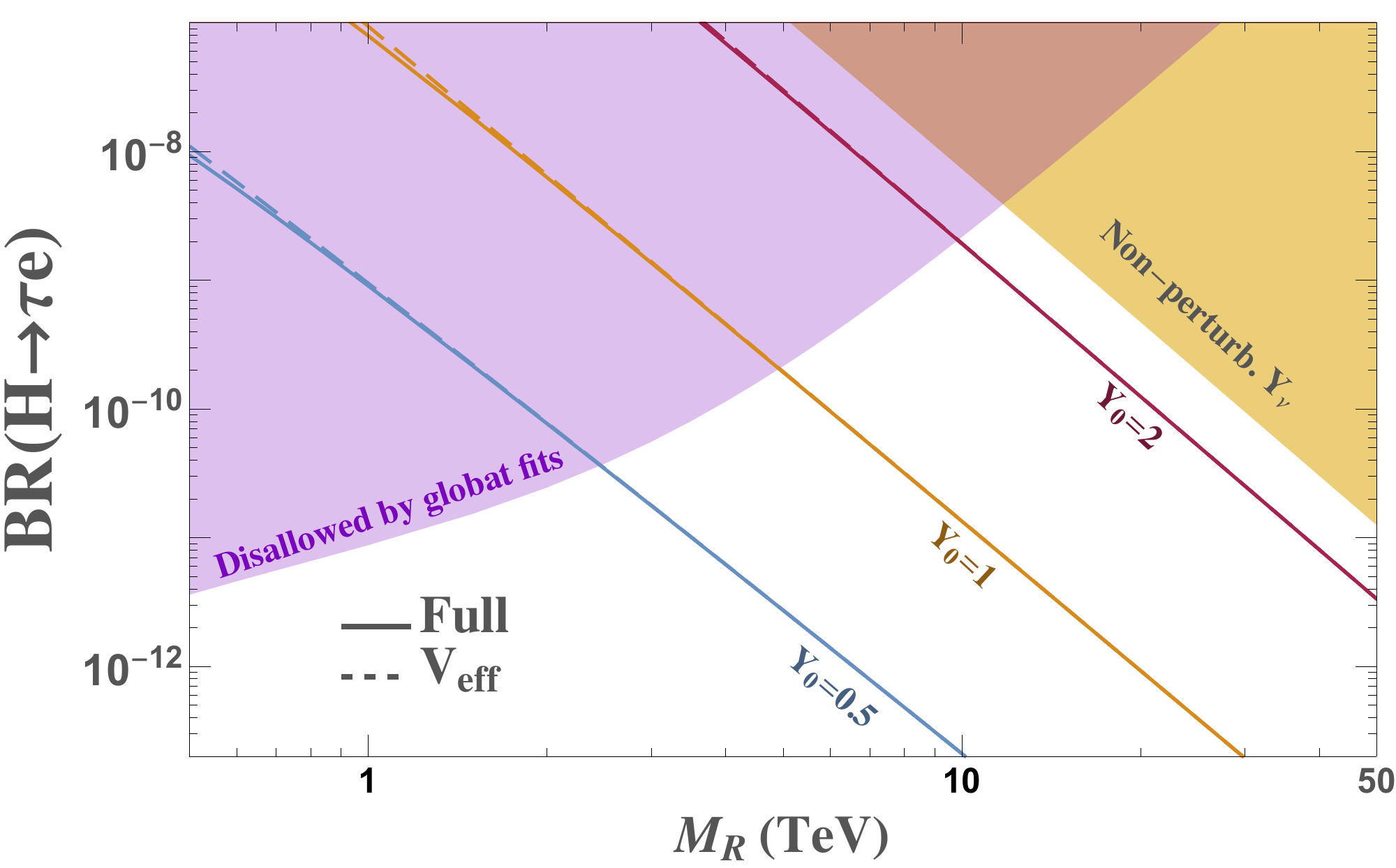}
\includegraphics[width=.49\textwidth]{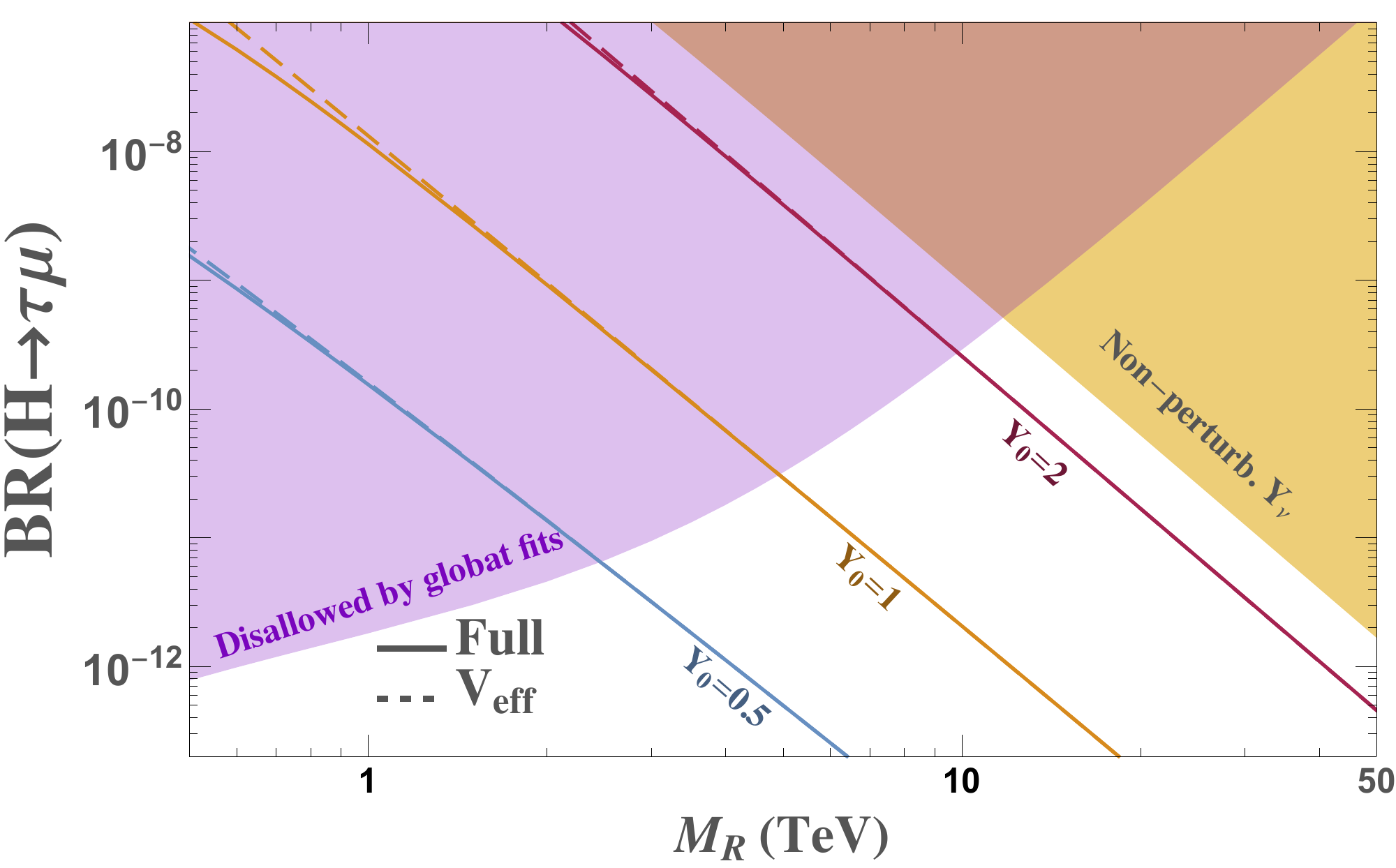}
\caption{Predictions for the Higgs boson decays into $\mu e$ (top), $\tau e$ (left) and $\tau\mu$ (right) computed with the effective vertex in the MIA (dashed lines) and the full results in the physical basis (solid lines) in the context of a low scale seesaw model with Yukawa coupling $\YLSS$ given by Eq.~\eqref{YLSSexample}. Shadowed areas are disallowed by global fits constraints (purple) and the non-perturbative regime for the Yukawa couplings (yellow), which we define as $|(Y_\nu Y_\nu^\dagger)^{km}|>4\pi$ for any element.}
\label{LFVHD_bounds}
\end{center}
\end{figure}

Let us consider a final example to illustrate this latter point. 
In order to avoid issues generating too large contributions to light neutrino masses, an elegant solution is to consider a low scale seesaw realization of the model, as we introduced in Eq.~\eqref{MassLSS}. 
If we assume, for example, that there are two pairs of sterile fermions ($n_\nu=2$) contributing to our observable and that they have the same mass, $M_{R_1}=M_{R_2}\equiv M_R$, an interesting value for the Yukawa matrix is the following:
\begin{equation}
\YLSS =Y_0 \left(\begin{array}{cc} 1 & 0 \\ 0.008 & 0.42 \\ 1 & 1.1 \end{array}\right)\,.
\label{YLSSexample}
\end{equation}
This example is useful as it leads to a $\eta$ matrix with a very similar pattern than that in Eq.~\eqref{etamax}.
Then, depending on the value of the global strength factor $Y_0$ and the mass $M_{R}$, we can define which part of the parameter space is allowed.

We show in Fig.~\ref{LFVHD_bounds} the predictions for the three LFVHD channels in this particular example, although we expect similar results for other models with more RH neutrinos as long as they lead to the similar $\eta$ matrices. 
The first thing we see from this figure is that the effective vertex (dashed lines) matches very well the full prediction (solid lines), as we already saw before.
Nevertheless, the simple expression of Eq.~\eqref{VeffGSS} allows us to easily understand the dependence on the different parameters of the model. 
Finally, we can also use this figure to deduce how large LFVHD rates can we expect after having considered the bounds in Eq.~\eqref{etamax} as well as perturbativity bounds.
As we discussed above, the largest possible values are obtained for heavy masses at around 10~TeV, when the bounds from the global fit analysis and perturbativity cross. 
Unfortunately, the branching ratios in the allowed white area are too small and far from current experimental sensitivities.


\section{Conclusions}
\label{sec:conclusions}

In this work, we have discussed the importance of having expressions for lepton flavor violating transitions which are expressed directly in terms of the fundamental parameters of the model. 
These expressions are helpful to better understand the observable, as well as to compare it with experimental observations in order to constrain the interaction parameter space.
In this context, the mass insertion approximation technique is a powerful tool, which we have reviewed with two simple examples. 

We have then studied  the LFV Higgs decays in the context of a general type-I seesaw model with an arbitrary number of right-handed neutrinos. 
We applied the MIA technique to this model, which allowed us to derived an effective vertex for $H\ell_k\ell_m$ after integrating out the heavy right-handed neutrinos. 
This analytical expression is useful to understand the behavior  of the observables with the fundamental parameters of the neutrino sector, i.e., the Yukawa couplings and the heavy Majorana masses.
Moreover, it also provides an alternative way to quickly evaluate these observables to a very good approximation, without the need of long numerical evaluations of the full result in the physical basis. 

Finally, we have made the connection to the phenomenologically interesting case of low scale seesaw models.
After explicitly checking that we recover existing results for the inverse seesaw case, we have evaluated the LFVHD rates taking into account current bounds from global analysis, as well as perturbativity bounds for the Yukawa couplings. 
Unfortunately, the predicted rates for the LFVHD in the allowed area are far from current experimental sensitivities and, consequently, they do not provide a competitive way of probing the existence of these heavy Majorana neutrinos.

\section*{Acknowledgments}
We would like to thank Claudia Garc\'ia-Garc\'ia for a careful reading of the manuscript.
This work is supported by the European Union through the ITN ELUSIVES H2020-MSCA-ITN-2015//674896 and the RISE INVISIBLESPLUS H2020-MSCA-RISE-2015//690575, by the CICYT through the project FPA2016-78645-P, and by the Spanish MINECO’s ``Centro de Excelencia Severo Ochoa'' Programme under grant SEV-2016-0597.

\begin{appendix}
\section{MIA Form factors}\label{AppA}

\begin{figure}[t!]
\begin{center}
\vspace{1cm}
\includegraphics[scale=.75]{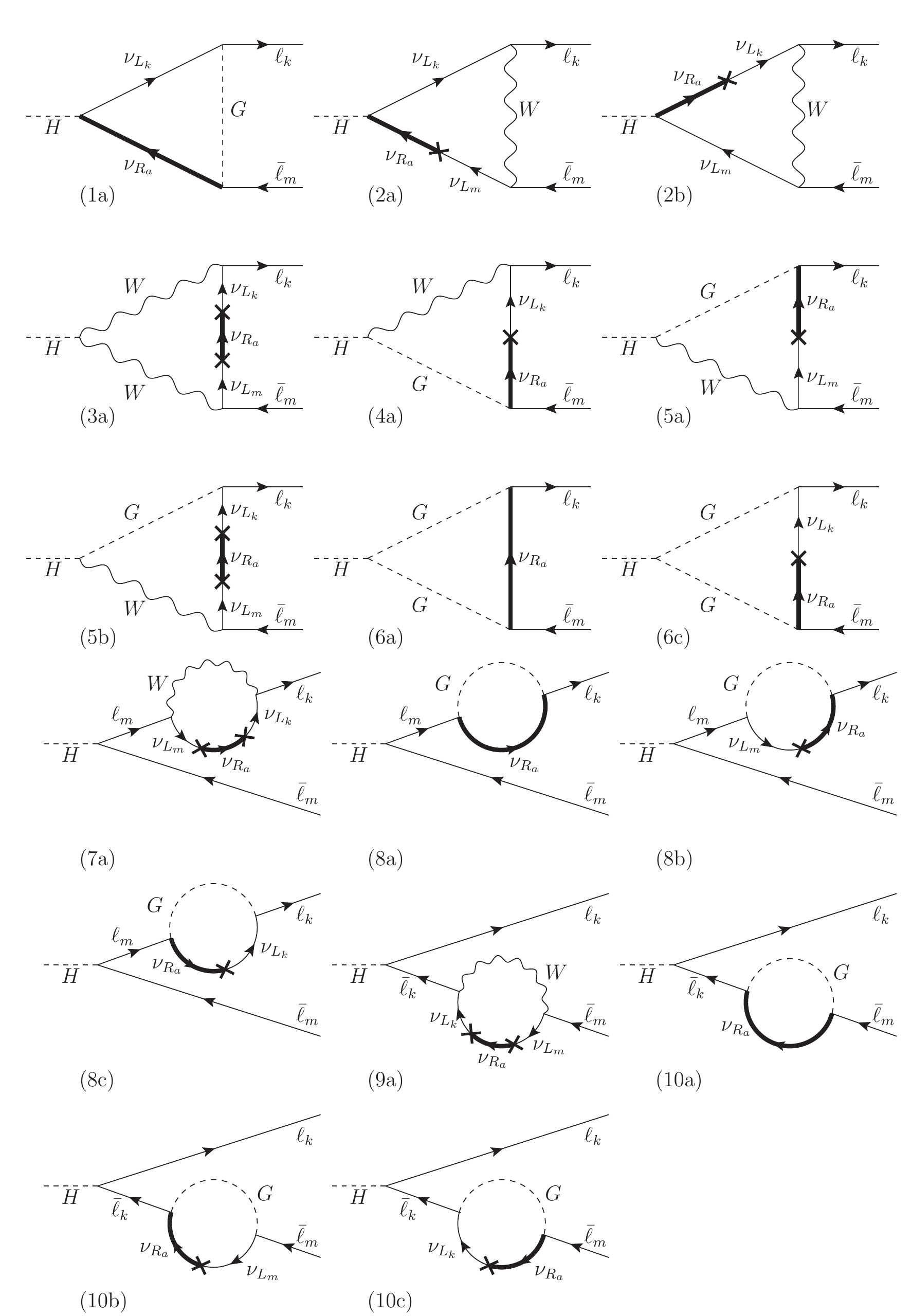}
\caption{Dominant diagrams contributing at ${\cal O} (Y_\nu^2)$ within the MIA.}\label{dominant-Y2}
\end{center}
\end{figure}
\begin{figure}[t!]
\begin{center}
\includegraphics[scale=.75]{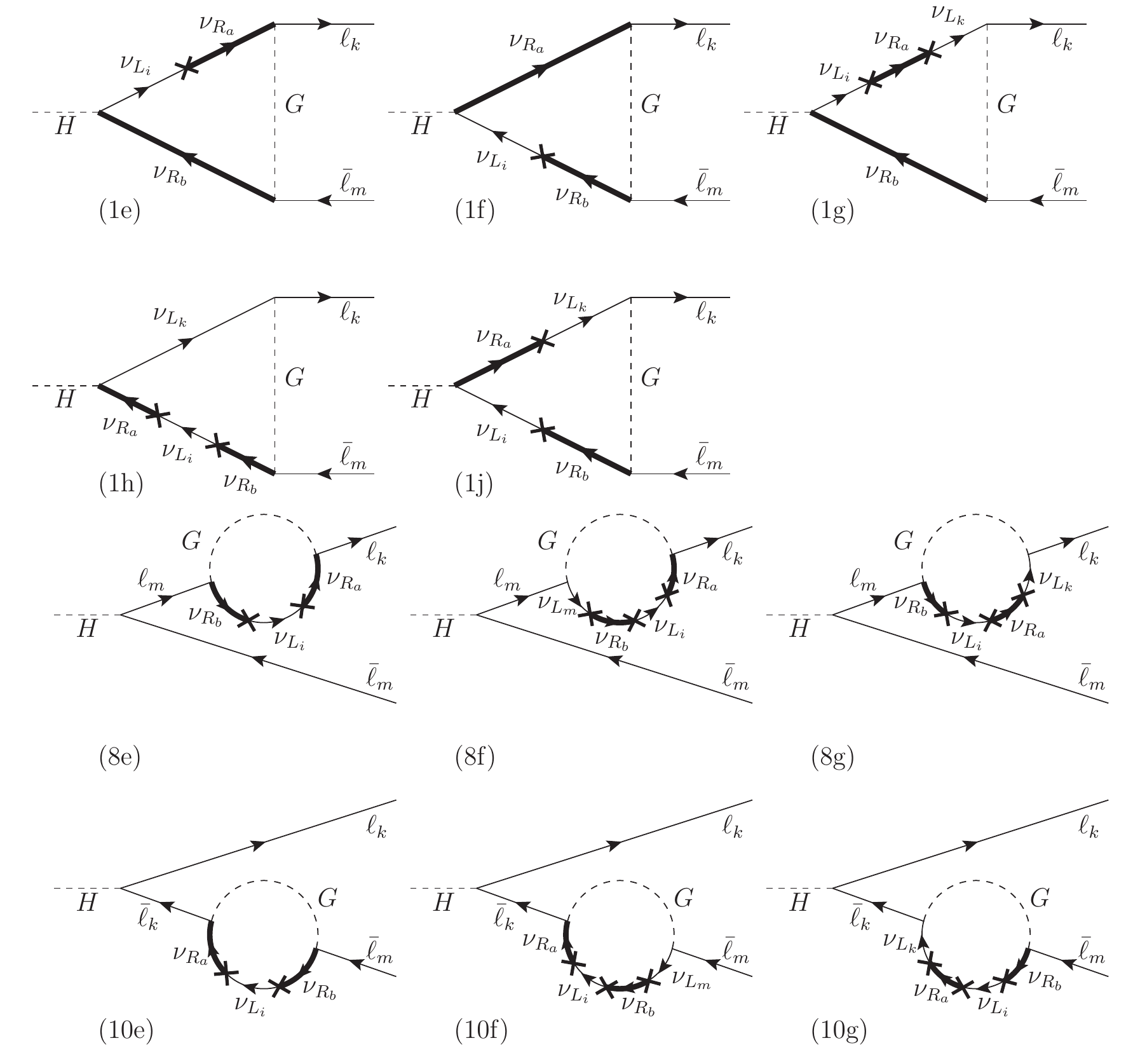}
\caption{Dominant diagrams contributing at ${\cal O} (Y_\nu^4)$ within the MIA.}\label{dominant-Y4}
\end{center}
\end{figure}

The analytical computation of the one-loop effective vertex in the MIA approach are written in terms of the one-loop functions for $D$ dimensions in the Passarino-Veltman notation~\cite{Passarino:1978jh}, following these definitions and conventions:
\begin{equation}\label{loopfunctionB}
    \mu^{4-D}~ \int \frac{d^D k}{(2\pi)^D} \frac{\{1; k^{\mu}\}}
    {[k^2 - m_1^2][(k+q_1)^2 - m_2^2]}
    = \frac{i}{16\pi^2} \left\{ B_0; q_1^{\mu} B_1 \right\}
    (q_1,m_1,m_2)\,,
\end{equation}
\begin{align}
    \mu^{4-D}& \int \frac{d^D k}{(2\pi)^D}
    \frac{\{1; k^2; k^{\mu}\}}
    {[k^2 - m_1^2][(k + q_1)^2 - m_2^2][(k + q_1 + q_2)^2 - m_3^2]}
    \nonumber \\
     & = \frac{i}{16\pi^2}
    \left\{ C_0; {\tilde C}_0; q_1^{\mu}C_{11} + q_2^{\mu}C_{12} \right\}
    (q_1, q_2, m_1, m_2, m_3)\,, \label{loopfunctionC}
\end{align}
\begin{align}
   \mu^{4-D}& \int \frac{d^D k}{(2\pi)^D}
    \frac{\{1; k^2; k^{\mu}\}}
    {[k^2 - m_1^2][(k + q_1)^2 - m_2^2][(k + q_1 + q_2)^2 - m_3^2][(k + q_1 + q_2+ q_3)^2 - m_4^2]}
    \nonumber \\
     & = \frac{i}{16\pi^2}
    \left\{ D_0; {\tilde D}_0; q_1^{\mu}D_{11} + q_2^{\mu}D_{12} +q_3^{\mu}D_{13}\right\}
    (q_1, q_2, q_3, m_1, m_2, m_3,m_4)\,. \label{loopfunctionD}
\end{align}

As explained in the text, we consider the hierarchy between charged lepton masses ${m_{\ell_m} \ll m_{\ell_k}}$ and thus we only consider the left-handed form factor $F_L$ in Eq.~\eqref{amplitude}. 
In order to simplify our expressions we neglect the contributions of ${\cal O}(m_{\ell}^3)$ in the form factors because they are very suppressed and we recovered the full predictions without them.
We present the analytical results for the left-handed form factors computed within the MIA to one-loop order and considering the leading order contributions, ${\cal O} (Y_\nu^2)$, and the next to leading order corrections, ${\cal O} (Y_\nu^4)$, as explained in the text. We collect in Figs.~\ref{dominant-Y2}-\ref{dominant-Y4}, the relevant diagrams in the Feynman-'t Hooft gauge. We follow the notation for the diagrams given in~\cite{Arganda:2017vdb}. Missing diagrams correspond to suppressed contributions proportional to higher lepton masses powers.

For shortness, we introduce the factor $\Ffactor=\frac{1}{32 \pi^{2}} \frac{m_{\ell_k}}{m_{W}}$, so that the results of the ${\cal O} (Y_\nu^2)$ dominant contributions diagram by diagram are: 
\begin{align}
 F_{L}^{{\rm MIA\, (1)}}\big|_{Y_\nu^2}=& \Ffactor  \left(Y_{\nu}^{ka} Y_{\nu}^{\dagger am}  \right) (\tilde{C}_{0})_{(1a)}  \,,\nonumber \\
F_{L}^{{\rm MIA\, (2)}}\big|_{Y_\nu^2}=& -2\Ffactor\,m_W^2 \left(Y_{\nu}^{ka} Y_{\nu}^{\dagger am} \right) \Big( (C_{0}+C_{11}-C_{12})_{(2a)} +(C_{11}-C_{12})_{(2b)} \Big) \,,\nonumber \\
 F_{L}^{{\rm MIA\, (3)}}\big|_{Y_\nu^2}=& 4\Ffactor\,m_{W}^{4} \left(Y_{\nu}^{ka} Y_{\nu}^{\dagger am} \right) ( D_{12}-D_{13} )_{(3a)} \,,\nonumber \\
 F_{L}^{{\rm MIA\, (4)}}\big|_{Y_\nu^2}=& -\Ffactor\,m_{W}^2 \left(Y_{\nu}^{ka} Y_{\nu}^{\dagger am} \right) (C_{0}-C_{11}+C_{12})_{(4a)}  \,, \nonumber \\
 F_{L}^{{\rm MIA\, (5)}}\big|_{Y_\nu^2}=& \Ffactor\,m_{W}^2  \left(Y_{\nu}^{ka} Y_{\nu}^{\dagger am}  \right) \Big( (2C_{0}+C_{11}-C_{12})_{(5a)} +(C_{0}-2m_{H}^{2}D_{13})_{(5b)} \Big) \,, \nonumber\\
 F_{L}^{{\rm MIA\, (6)}}\big|_{Y_\nu^2}=& \Ffactor\,m_{H}^{2} \left( Y_{\nu}^{ka} Y_{\nu}^{\dagger am} \right) \Big( (C_{11}-C_{12})_{(6a)} +(C_{0})_{(6c)} \Big) \,,\nonumber \\
 F_{L}^{{\rm MIA\, (7)}}\big|_{Y_\nu^2}=& 2\Ffactor\,m_W^2 \frac{m_{\ell_m}^{2}}{m_{\ell_k}^{2}-m_{\ell_m}^{2}}  \left(Y_{\nu}^{ka} Y_{\nu}^{\dagger am} \right) ( C_{12} )_{(7a)} \,, \nonumber \\
  F_{L}^{{\rm MIA\, (8)}}\big|_{Y_\nu^2}=& \Ffactor \frac{m_{\ell_m}^{2}}{m_{\ell_k}^{2}-m_{\ell_m}^{2}}  \left( Y_{\nu}^{ka} Y_{\nu}^{\dagger am} \right) \Big( (B_{1})_{(8a)} +(B_{0})_{(8b)} +(B_{0})_{(8c)}  \Big)\,, \nonumber \\
 F_{L}^{{\rm MIA\, (9)}}\big|_{Y_\nu^2}=& -2\Ffactor\,m_{W}^2 \frac{m_{\ell_m}^{2}}{m_{\ell_k}^{2}-m_{\ell_m}^{2}}  \left(Y_{\nu}^{ka} Y_{\nu}^{\dagger am} \right) ( C_{12} )_{(9a)}\,, \nonumber \\
 F_{L}^{{\rm MIA\, (10)}}\big|_{Y_\nu^2}=& -\Ffactor \frac{m_{\ell_m}^{2}}{m_{\ell_k}^{2}-m_{\ell_m}^{2}}  \left( Y_{\nu}^{ka} Y_{\nu}^{\dagger am} \right) \Big( (B_{1})_{(10a)} +(B_{0})_{(10b)} +\frac{m_{\ell_k}^{2}}{m_{\ell_m}^{2}}(B_{0})_{(10c)} \Big) \, .
\label{FLtot_op2_Y2}
\end{align}
On the other hand, the results of the dominant ${\cal O} (Y_\nu^4)$ contributions are:
\begin{align}
F_{L}^{{\rm MIA\, (1)}}\big|_{Y_\nu^4}=&  \Ffactor\,v^2
\left(Y_{\nu}^{ka} Y_{\nu}^{\dagger ai} Y_{\nu}^{ib} Y_{\nu}^{\dagger bm} \right) \Big( -(C_{11}-C_{12})_{(1e)} -(C_{11}-C_{12}+C_{0})_{(1f)}    \nonumber \\
&+(\tilde{D}_{0})_{(1g)} +(\tilde{D}_{0})_{(1h)} +(C_{0})_{(1j)} \Big)\,,
\nonumber \\
 F_{L}^{{\rm MIA\, (8)}}\big|_{Y_\nu^4}=& \Ffactor\,v^2 \frac{m_{\ell_m}^{2}}{m_{\ell_k}^{2}-m_{\ell_m}^{2}}
 \left(Y_{\nu}^{ka} Y_{\nu}^{\dagger ai} Y_{\nu}^{ib} Y_{\nu}^{\dagger bm}  \right) \Big( (C_{12})_{(8e)}+(C_{0})_{(8f)}+(C_{0})_{(8g)} \Big) \,,  \nonumber \\
F_{L}^{{\rm MIA\, (10)}}\big|_{Y_\nu^4}=& -\Ffactor\,v^2 \frac{m_{\ell_m}^{2}}{m_{\ell_k}^{2}-m_{\ell_m}^{2}}
\left(Y_{\nu}^{ka} Y_{\nu}^{\dagger ai} Y_{\nu}^{ib} Y_{\nu}^{\dagger bm}  \right) \Big( (C_{12})_{(10e)}+(C_{0})_{(10f)}+\frac{m_{\ell_k}^{2}}{m_{\ell_m}^{2}}(C_{0})_{(10g)} \Big) \, .  
\label{FLtot_op2_Y4}
\end{align}
In the above expressions, an implicit sum over $i=1\dots3$ and $a,b=1\dots N$ has to be understood, and the arguments of the one-loop integrals are: 
\begin{align}
 \tilde{C}_{0}, C_{i} =& \tilde{C}_{0}, C_{i} (p_{2},p_{1},m_{W},0,M_{a}) && \text{in } (1a), (2a) \,,\nonumber \\
 \tilde{C}_{0}, C_{i} =& \tilde{C}_{0}, C_{i} (p_{2},p_{1},m_{W},M_{a},0) && \text{in } (2b) \,, \nonumber \\
 C_{i} =& C_{i} (p_{2},p_{1},m_{W},M_{a},M_{b}) && \text{in } (1e), (1f), (1j) \,, \nonumber \\
 \tilde{D}_{0} =& \tilde{D}_{0} (p_{2},0,p_{1},m_{W},0,M_{a},M_{b}) && \text{in } (1g) \,, \nonumber \\
 \tilde{D}_{0} =& \tilde{D}_{0} (p_{2},p_{1},0,m_{W},0,M_{a},M_{b}) && \text{in } (1h) \,, \nonumber \\
 D_{i} =& D_{i} (0,p_{2},p_{1},0,M_{a},m_{W},m_{W}) && \text{in } (3a), (5b) \,, \nonumber\\
 C_{i} =&  C_{i} (p_{2},p_{1},M_{a},m_{W},m_{W}) && \text{in } (4a), (5a), (5b), (6a), (6c) \,, \nonumber \\
 C_{12} =& C_{12} (0,p_{2},0,M_{a},m_{W}) && \text{in } (7a)  \,,\nonumber \\
 B_{i} =& B_{i} (p_{2},M_{a},m_{W}) && \text{in } (8a), (8b), (8c)  \,,\nonumber \\
 C_{i} =& C_{i} (0,p_{2},M_{a},M_{b},m_{W}) && \text{in } (8e), (8f),(8g)  \,, \nonumber \\
 C_{12} =& C_{12} (0,p_{3},0,M_{a},m_{W}) && \text{in } (9a) \,, \nonumber \\
 B_{i} =& B_{i} (p_{3},M_{a},m_{W}) && \text{in } (10a), (10b), (10c) \,, \nonumber \\
 C_{i} =& C_{i} (0,p_{3},M_{a},M_{b},m_{W}) && \text{in } (10e), (10f),(10g) \,.
\label{argfloops_op2}
\end{align}

\section{Heavy mass expansions}
\label{AppB}
\end{appendix}

In order to get analytical expressions for the left-handed form factors of the previous Appendix, we simplify the corresponding one-loop functions by means of the heavy Majorana masses involved in the loops.  
We performed the integrals of Eqs.~(\ref{loopfunctionB}-\ref{loopfunctionD}), for the cases involved in the present LFVHD calculation, with the Feynman parameters and using the kinematical relations coming from the momentum conservation $p_1=p_2-p_3$ and considering the external particles on-shell:
\ba
p_1^2 = m_H^2\,,& \quad &p_2 \cdot p_3 = \frac{1}{2} \left( -m_H^2 +m_{\ell_k}^2 +m_{\ell_m}^2 \right) \approx -\frac{m_H^2}{2}\,,  \nonumber\\
p_2^2 = m_{\ell_k}^2 \approx 0\,,& \quad &p_3 \cdot p_1 = \frac{1}{2} \left( m_H^2 -m_{\ell_k}^2 +m_{\ell_m}^2 \right) \approx \frac{m_H^2}{2}\,,  \nonumber\\
p_3^2 = m_{\ell_m}^2 \approx 0\,,& \quad &p_1 \cdot p_2 = \frac{1}{2} \left( -m_H^2 -m_{\ell_k}^2 +m_{\ell_m}^2 \right) \approx -\frac{m_H^2}{2}\,.
\label{kinematics}
\ea

At this step, each integral is written in terms of the lepton, $W$ and Higgs bosons and the Majorana masses. We need an approximation in order to perform these integrals analytically: we assume the following hierarchy $m_{\ell_m}, m_{\ell_k}\ll m_W, m_H, v\ll M$ between the three mass scales participating in the calculation, i.e., lepton masses, EW scale and the new physics scale(s). Therefore, we neglect the lepton masses in the integrals. We are interested in the dominant contributions of the large Majorana mass limit, i.e., we keep the corrections of $\mathcal{O}(m_W^2/M^2)$, $\mathcal{O}(m_H^2/M^2)$ and $\mathcal{O}(v^2/M^2)$ (the logarithmic dependences are not expanded). 

Using the standard definitions in dimensional regularization, $\Delta= 2/\epsilon-\gamma_E +{\rm Log}(4\pi)$ with $D=4-\epsilon$ and $\mu$ the usual scale, and defining the mass ratio $\lambda=\frac{m_{W}^{2}}{m_{H}^{2}}$.
We find the following expressions for the dominant contributions of all the one-loop functions concerned in this work:
\ba
B_{0} \left(q,M,m_W\right) &=& \Delta +1 -\log \Big(\frac{M^{2}}{\mu^2}\Big) +\frac{m_{W}^{2} \log \left(\frac{m_{W}^{2}}{M^{2}}\right)}{M^{2}} +\frac{q^2}{2M^2}  \,,\nonumber\\
C_{0} \left(p_2,p_1,m_W,0,M\right) &=& C_{0} \left(p_2,p_1,m_W,M,0\right) = \frac{\log \left(\frac{m_{W}^{2}}{M^{2}}\right)}{M^{2}}  \,, \nonumber\\
C_{0} \left(p_2,p_1,M,m_W,m_W\right) &=& \frac{2 \sqrt{4 \lambda-1} \arctan \left(\sqrt{\frac{1}{4 \lambda-1}}\right)-1+\log \left(\frac{m_{W}^{2}}{M^{2}}\right)}{M^{2}} \,,  \nonumber\\
C_{0} \left(p_2,p_1,m_W,M_a,M_b\right)&=& C_{0} \left(p_2,p_1,M_{a},M_{b},m_W\right) = -\frac{\log \left(\frac{M_{a}^{2}}{M_{b}^{2}}\right)}{M_{a}^{2}-M_{b}^{2}} \,, \nonumber\\
\tilde{C}_{0} \left(p_2,p_1,m_W,M,0\right) &=& \tilde{C}_{0} \left(p_2,p_1,m_W,0,M\right) \nonumber\\
&=& \Delta 
+1 -\log \Big(\frac{M^{2}}{\mu^2}\Big)  + \frac{m_{W}^{2} \log \left(\frac{m_{W}^{2}}{M^{2}}\right)}{M^{2}}+\frac{m_{H}^{2}}{2 M^{2}} \,,  \nonumber\\
\tilde{D}_{0} \left(p_2,0,p_1,m_W,0,M_{a},M_{b}\right) &=& \tilde{D}_{0} \left(p_2,p_1,0,m_W,0,M_{a},M_{b}\right) =  -\frac{\log \left(\frac{M_{a}^{2}}{M_{b}^{2}}\right)}{M_{a}^{2}-M_{b}^{2}} \,,\nonumber\\
\tilde{D}_{0} \left(p_2,0,p_1,m_W,M_{a},M_{b},0\right) &=& \tilde{D}_{0} \left(p_2,p_1,0,m_W,M_{a},M_{b},0\right) = -\frac{\log \left(\frac{M_{a}^{2}}{M_{b}^{2}}\right)}{M_{a}^{2}-M_{b}^{2}} \,, \nonumber\\
B_{1} \left(q,M,m_W\right) &=& -\frac{\Delta }{2}-\frac{3}{4}+\frac12\log\Big(\frac{M^{2}}{\mu^2}\Big)-\frac{m_{W}^{2} \left(2 \log \left(\frac{m_{W}^{2}}{M^{2}}\right)+1\right)}{2 M^{2}}  -\frac{q^2}{3M^2}\,, \nonumber\\
C_{11} \left(p_2,p_1,m_W,0,M\right) &=& \frac{1-\log \left(\frac{m_{W}^{2}}{M^{2}}\right)}{2 M^{2}}  \,, \nonumber\\
C_{12} \left(p_2,p_1,m_W,0,M\right) &=& \frac{1}{2 M^{2}}  \,, \nonumber\\
C_{11} \left(p_2,p_1,m_W,M,0\right) &=& \frac{1-\log \left(\frac{m_{W}^{2}}{M^{2}}\right)}{2 M^{2}}   \,,\nonumber\\
C_{12} \left(p_2,p_1,m_W,M,0\right) &=& -\frac{\log \left(\frac{m_{W}^{2}}{M^{2}}\right)}{2 M^{2}}   \,,\nonumber\\
C_{11} \left(p_2,p_1,M,m_W,m_W\right) &=& 2 C_{12} \left(p_2,p_1,M,m_W,m_W\right)  \nonumber\\
&=& -\frac{4 \sqrt{4 \lambda-1} \arctan \left(\sqrt{\frac{1}{4 \lambda-1}}\right)+2 \log \left(\frac{m_{W}^{2}}{M^{2}}\right)-1}{2 M^{2}} \,,  \nonumber\\
C_{11} \left(p_2,p_1,m_W,M_a,M_b\right) &=& 
\frac{\log \left(\frac{M_{a}^{2}}{M_{b}^{2}}\right)}{2(M_{a}^{2}-M_{b}^{2})} \,,
 \nonumber\\
  C_{12} \left(p_2,p_1,m_W,M_a,M_b\right) &=&
 \frac{M_{b}^{2}-M_{a}^{2}+M_{a}^{2}\log \left(\frac{M_{a}^{2}}{M_{b}^{2}}\right)}{2 (M_{a}^{2}-M_{b}^{2})^2} \,,
 \nonumber\\
 C_{12} \left(p_2,p_1,M_a,M_b,m_W\right) &=&  
\frac{\log \left(\frac{M_{a}^{2}}{M_{b}^{2}}\right)}{2(M_{a}^{2}-M_{b}^{2})} \,,
 \nonumber\\
C_{12} \left(p_2,p_1,0,M,m_W\right)&=& \frac{-\log \left(\frac{m_{W}^{2}}{M^{2}}\right)-1}{2 M^{2}}  \,,\nonumber\\
D_{12} \left(0,p_2,p_1,0,M,m_W,m_W\right) &=& 2 D_{13} \left(0,p_2,p_1,0,M,m_W,m_W\right)   \nonumber\\
&=& \frac{2\left( -4 \lambda \arctan^2 \left(\sqrt{\frac{1}{4 \lambda-1}}\right)+2 \sqrt{4 \lambda-1} \arctan \left(\sqrt{\frac{1}{4 \lambda-1}}\right)-1 \right)}{ M^{2} m_{H}^{2}} \, . \nonumber\\
\label{allfloops}
\ea
Although we have derived these expressions assuming the approximations in Eq.~\eqref{kinematics}, we have checked that they are valid as long as  $p_2^2=0$ and in the branch $p_1^2<4 m_W^2$.

We checked that the above expansions are in very good agreement with the numerical results from LoopTools~\cite{Hahn:1998yk}. 
We also checked that the one-loop functions with two different Majorana masses $M_a$ and $M_b$ reproduce the corresponding expressions in the degenerate limit $M_a=M_b=M_R$ given in~\cite{Arganda:2017vdb}.

An important comment is in order: an usual approximation to do those type of integrals is the zero external momentum ($p_1^2=p_2^2=p_3^2=0$). If we apply this approximation, the Higgs boson mass coming from Eq.~\eqref{kinematics} is missed at the beginning. In the present LFVHD computation, this fact provides wrong predictions due to the presence of the $W$ boson mass in the loop: the corrections of $\mathcal{O}(m_H^2/M^2)$ and $\mathcal{O}(m_W^2/M^2)$ are of the same order. 
 For example, for the $C_0$ one-loop function involved in diagrams of type (4), (5) and (6) we have the dominant contributions of the large Majorana mass limit in the zero external momentum approximation:
\ba
C_0(p_2^2=0,p_1^2=0,M,m_W,m_W) &=& \frac{1 +\log \left( \frac{m_W^2}{M^2} \right)}{M^2} \,,
\ea
Although this equation corresponds to the third line of Eq.~\eqref{allfloops} in the limit $p_1^2\to0$, it is not enough to reproduce the full result for the LFVHD. The reason is the fact that the contributions of the on-shell Higgs boson momentum are competitive with the $W$ boson mass in the loop.

Finally, the LO and NLO contributions of the left-handed form factors (grouped in convenient pairs) in the large Majorana mass limit are:
\ba
F_{L}^{{\rm MIA\, (1)}} &\approx& \Ffactor \Bigg[ \left(Y_{\nu}^{ka} Y_{\nu}^{\dagger am}\right) \left( \Delta +1 -\log \left(\frac{M_{a}^{2}}{\mu^2}\right) + \frac{m_{W}^{2} \log \left(\frac{m_{W}^{2}}{M_{a}^{2}}\right)}{M_{a}^{2}}+\frac{m_{H}^{2}}{2 M_{a}^{2}} \right)   \nonumber\\
&& \hspace{8mm}-v^2\left(Y_{\nu}^{ka} (Y_{\nu}^{\dagger} Y_{\nu})^{ab} Y_{\nu}^{\dagger bm} \right)\frac{M_a^2-M_b^2+(2M_a^2-3M_b^2)\log \left(\frac{M_a^{2}}{M_{b}^{2}}\right)}{(M_a^2-M_b^2)^2}  \Bigg] \,,\nonumber\\
F_{L}^{{\rm MIA\, (2)}} &\approx& -\Ffactor  \left(Y_{\nu}^{ka} Y_{\nu}^{\dagger am}\right) \frac{m_{W}^{2}}{M_{a}^{2}} \left( 1+\log \left(\frac{m_{W}^{2}}{M_{a}^{2}}\right) \right)\,,   \nonumber\\
F_{L}^{{\rm MIA\, (3)}} &\approx& 4\Ffactor \left(Y_{\nu}^{ka} Y_{\nu}^{\dagger am} \right) \frac{\lambda m_{W}^{2}}{M_{a}^{2}} \left( -4\lambda \arctan^2 \left(\frac{1}{\sqrt{4\lambda-1}}\right) \right. \nonumber\\
&& \left. +2\sqrt{4\lambda-1}\arctan \left(\frac{1}{\sqrt{4\lambda-1}}\right) -1  \right) \,,\nonumber\\
F_{L}^{{\rm MIA\, (4+5)}} &\approx& \Ffactor \left(Y_{\nu}^{ka} Y_{\nu}^{\dagger am} \right) \frac{m_{W}^{2}}{M_{a}^{2}} \left( 8\lambda \arctan^2 \left(\frac{1}{\sqrt{4\lambda-1}}\right) \right. \nonumber\\
&& \left. -2\sqrt{4\lambda-1}\arctan \left(\frac{1}{\sqrt{4\lambda-1}}\right) +\frac{1}{2} +\log \left(\frac{m_{W}^{2}}{M_{a}^{2}}\right)  \right)\,, \nonumber\\
F_{L}^{{\rm MIA\, (6)}} &\approx& \Ffactor \left(Y_{\nu}^{ka} Y_{\nu}^{\dagger am} \right) \frac{m_{H}^{2}}{M_{a}^{2}} \left( \sqrt{4\lambda-1}\arctan \left(\frac{1}{\sqrt{4\lambda-1}}\right) -\frac{3}{4} +\frac{\log \left(\frac{m_{W}^{2}}{M_{a}^{2}}\right)}{2}  \right) \,, \nonumber\\
F_{L}^{{\rm MIA\, (7+9)}} &\approx& 0 \,,\nonumber\\
F_{L}^{{\rm MIA\, (8+10)}} &\approx& -\Ffactor \Bigg[ \left(Y_{\nu}^{ka} Y_{\nu}^{\dagger am}\right) \left( \Delta +1 -
\log \left(\frac{M_{a}^{2}}{\mu^2}\right) + \frac{m_{W}^{2} \log \left(\frac{m_{W}^{2}}{M_{a}^{2}}\right)}{M_{a}^{2}} \right)   \nonumber\\
&&\hspace{11mm} -v^2\left(Y_{\nu}^{ka} (Y_{\nu}^{\dagger} Y_{\nu})^{ab} Y_{\nu}^{\dagger bm} \right)\frac{\log \left(\frac{M_{a}^{2}}{M_{b}^{2}}\right)}{M_{a}^{2}-M_{b}^{2}}  \Bigg] \,,  
\label{FLsimple_totdom}
\ea
where the index $i$ corresponding to the left-handed neutrinos was contracted in the $Y_\nu^4$ contributions.


\bibliography{biblio}

\end{document}